# Multi-agent Modeling of Hazard-Household-Infrastructure Nexus for Equitable Resilience Assessment

Amir Esmalian[1] | Wanqiu Wang[2] | Ali Mostafavi[1]

[1]Zachry Department of Civil Engineering, Texas A&M University, College Station, TX, USA

[2] Department of Computer Science and Engineering, Texas A&M University, College Station, TX, USA

**Correspondence**
Ali Mostafavi, Zachry Department of Civil Engineering, Texas A&M University, College Station, TX 77840, USA
Email: amostafavi@civil.tamu.edu

**Funding information**
National Science Foundation, CAREER, Grant/Award Number: 1846069 and National Academies' Gulf Research Program Early-Career Research Fellowship

**ABSTRACT**

Infrastructure service disruptions impact households in an affected community disproportionately. To enable integrating social equity considerations in infrastructure resilience assessments, this study created a new computational multi-agent simulation model which enables integrated assessment of hazard, infrastructure system, and household elements and their interactions. With a focus on hurricane-induced power outages, the model consists of three elements: 1) the hazard component simulates exposure of the community to a hurricane with varying intensity levels; 2) the physical infrastructure component simulates the power network and its probabilistic failures and restoration under different hazard scenarios; and 3) the households component captures the dynamic processes related to preparation, information seeking, and response actions of households facing hurricane-induced power outages. We used empirical data from household surveys from three hurricanes (Harvey, Florence, and Michael) in conjunction with theoretical decision-making models to abstract and simulate the underlying mechanisms affecting experienced hardship of households when facing power outages. The multi-agent simulation model was then tested in the context of Harris County, Texas, and verified and validated using empirical results from Hurricane Harvey in 2017. Then, the model was used to examine effects of different factors—such as forewarning durations, social network types, and restoration and resource allocation strategies—on reducing the societal impacts of service disruptions in an equitable manner. The results show that improving the restoration prioritization strategy to focus on vulnerable populations is an effective approach, especially during high-intensity events, to enhance equitable resilience. The results show the capability of the proposed computational model for capturing the dynamic and complex interactions in the nexus of humans, hazards, and infrastructure systems to better integrate human-centric aspects in resilience planning and into assessment of infrastructure systems in disasters. Hence, the proposed model and its results could provide a new tool for infrastructure managers and operators, as well as for disaster managers, in devising hazard mitigation and response strategies to reduce the societal impacts of power outages in an equitable manner.

## 1 INTRODUCTION

The objective of this study is to create a computational multi-agent simulation framework for capturing dynamic processes and interactions in the nexus of hazards, humans, and infrastructure systems in order to better integrate social impacts and equality considerations in infrastructure resilience assessments. The societal impacts of prolonged disruptions in infrastructure systems are the emergent properties arising from dynamic interactions in complex socio-physical systems (Dai et al. 2020; Guidotti et al. 2019; Rasoulkhani et al. 2020; Williams et al. 2020). Therefore, there is a need for novel computational models to capture and model the dynamic processes and interactions between the complex systems of humans, hazards, and infrastructure systems. With a focus on prolonged power outages during hurricanes, this study proposes a novel simulation modeling framework for enabling the integration of hazard, household, and infrastructure systems to examine the societal impacts on infrastructure service disruptions.

Existing infrastructure resilience assessment models focus primarily on physical infrastructure but fall short of fully considering interactions between households and hazards and infrastructure (Mostafavi 2018; Mostafavi and Ganapati 2019). Computational frameworks properly model the failure and restoration of infrastructure systems in the face of disturbances to the systems (Guikema et al. 2014; Ouyang and Dueñas-Osorio 2014; Ouyang and Fang 2017; Tomar and Burton 2021; Winkler et al. 2010). Several studies have devised ways to assess the resilience of infrastructure systems (Batouli and Mostafavi 2018; Gori et al. 2020; Guidotti et al. 2019; Hassan and Mahmoud 2021; Ma et al. 2019). These studies include computational models for determining the system's reliability when exposed to potential hazards with respect to topological and inherent vulnerabilities (Figueroa-candia et al. 2018; Holmgren 2006; Mensah and Dueñas-Osorio 2016; Outages and Walsh 2018; Ouyang and Zhao 2014; Reed et al. 2010). Furthermore, there are frameworks which enable modeling and optimizing restoration of damaged infrastructure systems (Sharma et al. 2020; Sun and Davison 2019; Xu et al. 2019). While these studies inform about the resilience and reliability of physical infrastructure systems (such as power networks and transportation systems), shed light on the interactions between hazards and infrastructure, and include restoration time of utilities, the current body of literature lacks integrated computational models and frameworks that consider households' interactions with infrastructure systems vis-à-vis the probabilistic impacts of hazards.

Household-level attributes (e.g., previous hazard experience and socio-demographic attributes) and protective actions (e.g., preparedness and information seeking) and their integration with hazard scenarios, as well as consideration of probabilistic physical infrastructure failures, service disruption duration, and restoration possibilities, are essential components for examining societal impacts of infrastructure service disruptions. Recent research has shown a significant disparity in the societal impacts of infrastructure service disruptions (Chakalian et al. 2019; Coleman et al. 2019; Esmalian et al. 2020b; Mitsova et al. 2018, 2021). These studies unveil risk disparities and suggest that households are heterogeneous entities as evidenced by varying levels of tolerance for service disruptions. Particularly, shelter-in-place households experience great hardship from infrastructure service disruptions. A household's decisions related to protective actions are not only influenced by its attributes, such as socio-demographic characteristics, but they are also highly influenced by perceived risk from the hazard (Lindell and Hwang 2008), information-seeking process (Morss et al. 2016), and their social network's influence (Haer et al. 2016; Kashani et al. 2019). Capturing these dynamic processes and decisions is essential for modeling and understanding the societal impacts of infrastructure service disruptions. In addition, a households' hardship experiences are influenced largely by the duration of service disruptions, which is the result of physical infrastructure failures and the utilities' decisions regarding service restorations. Hence, the societal impacts of infrastructure service disruptions emerge from the complex interactions among various processes in the hazard, humans, and infrastructure systems nexus. The current literature, however, lacks computational models that are capable of capturing and modeling the complex interactions in this nexus.

To address this gap, this study proposes and tests a novel computational multi-agent simulation framework including three components: 1) the hazard component simulates exposure of the community to a hurricane with varying intensity levels; 2) the physical infrastructure component simulates the infrastructure network and its probabilistic failures and restoration processes under different hazard and resource allocation scenarios; and 3) the households component captures the dynamic processes related to preparation, information seeking, and response actions of households facing hurricane-induced service disruptions. The proposed modeling framework was tested in examination of strategies to reduce the societal impacts of disruptions of power systems. The model bridges the gap in the abstraction of behaviors of system components and provides computational implementation of households' interaction with infrastructure systems and probabilistic simulation of hazards, and failure scenarios to enable examining equitable ways for reducing the societal risks.

Using the proposed multi-agent computational simulation framework, we examined strategies to reduce the societal impacts of power outages and investigated important questions such as (1) What are the proper strategies for mitigating the societal risks due to prolonged power outages? (2) To what extent are the hazard mitigation and response interventions equitable? The model enables exploratory analysis of the pathways that determine levels of societal impacts. The model also enables assessing the extent to which different strategies for reducing the societal impacts are equitable (Williams et al. 2020). The computational modeling framework would help disaster managers, infrastructure managers, and utility operators in making informed decisions which consider the specific needs and societal risks in their resilience assessments.

## 2 MULTI-AGENT SIMULATION FRAMEWORK

We created a multi-agent simulation model which integrates physical and social resilience assessment to find strategies for reducing the societal impacts of prolonged power outages. Multi-agent simulation modeling is a proven approach for complex modeling and analysis of coupled human-infrastructure systems (Eid and El-adaway 2018; Nejat and Damnjanovic 2012; Rasoulkhani et al. 2020; Terzi et al. 2019). Multi-agent simulation model enables the consideration of dynamic processes and complex interactions among different entities (Haer et al. 2017; Watts et al. 2019; Widener et al. 2013). Furthermore, multi-agent simulation approach has the advantage of enabling the consideration of interrelation within agents and their heterogeneity (Morss et al. 2017; Navarrete Gutiérrez et al. 2017). Therefore, multi-agent simulation provides a powerful approach for modeling the nexus of hazard-human-infrastructure. In the context of this study, the hazard component would cause damage to the infrastructure systems and also influence the preparation time for households. The infrastructure system would be damaged due to natural hazards. The system's physical vulnerability and restoration decisions affect the duration of service outages. The experienced hardship due to service disruptions by individual households is a function of their susceptibility and protective actions. The susceptibility and protective actions of households are influenced by various factors (e.g., income and race) and processes and shape the level of tolerance of households to durations of service outages. Households perceive threats from the hazard, inform their social network, and make decisions about their protective actions (such as preparedness). Households in the community have unique attributes and interact with each other to inform their decision to take protective actions depending on their capabilities, perception of risks, and their immediate social network's actions. Thus, the dynamic process of information-seeking behavior and decision-making about the protective actions are integral aspects of determining the level of tolerance to power outages. When the duration of service outages exceeds the tolerance level of households, they would experience hardship (which is the indicator of societal impact in this study).

## 2.1 Model Overview

Figure 1 depicts the underlying mechanisms and processes in the hazard-humans-infrastructure nexus captured in the proposed framework. In this framework, each of the underlying mechanisms leading to the societal impacts could be captured as dynamic processes. The integration of these processes enables simulating the extent of infrastructure failures, tolerance level of households, and service restoration duration, and hence determines the proportion of households in the community which experience hardship under different scenarios of hazard intensity and response/restoration strategies.

The hazard component simulates the intensity of hazard and exposure of components of infrastructure systems. The infrastructure component captures the physical vulnerability and network topology of power infrastructure systems. The extent of damage to the infrastructure system depends on the components' fragility and the network topology. The more fragile the systems' components, the greater the probability of severe damage. Furthermore, network topology influences the system's physical vulnerability due to the cascading failure and connectivity loss in the network. The extent of damage and the restoration process of the utility determines the duration of service outage. The duration of power service outages effects the hardship experienced by households. (Miles and Chang 2011).

The household component captures the dynamic processes and interactions influencing the level of tolerance of households to service outages. In particular, this research focused on the shelter-in-place-households, as these households are vulnerable to the impacts of power outages. The rapidity of the unfolding of a hazard event affects how far in advance households are informed about the upcoming hazard event (i.e., hurricane), allowing them to take adequate protective actions. Households interact with each other to share information about the hurricane, as well as the potential duration of the outages based on the information they receive and characteristics specific to the household, such as prior hazard experience. Households make decisions about their protective action to reduce the impacts of service losses. Their decisions are not solely influenced by their risk perception and socio-demographic attributes; they are also influenced by other households' decisions. A household is more likely to prepare for an upcoming hurricane if another household in their social network prepares. Hence, the model captures the dynamic process related to the households' information search behavior, risk perception, and decisions related to preparedness actions that determine their tolerance. The experienced hardship of households would be determined by comparing their tolerance with their experienced duration of disruptions. The model could then simulate the hardship profile of the affected communities to examine societal impacts of varying hurricane intensities based on the physical condition of the power network, restoration activities, and households' protective actions to better tolerate the disruptions.

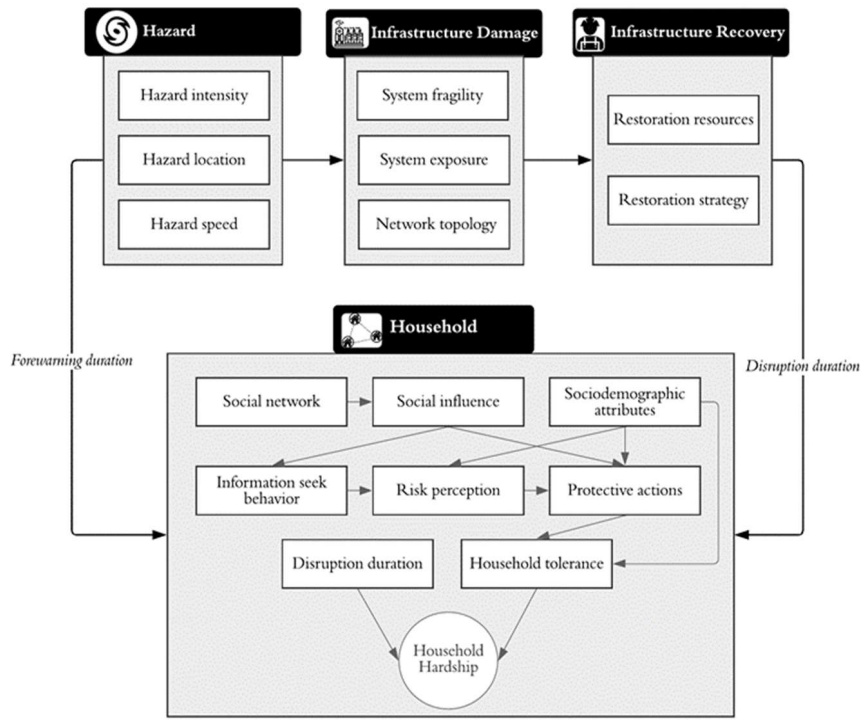

**FIGURE 1.** Human-hazard-infrastructure nexus framework for equitable resilience assessment of power systems.

## 2.2 Hazard agent

The hazard component of the proposed model considers failure of the power network due to damage by severe windstorms to components not designed to withstand strong winds. It is important to mention that the damages to the components of power network are not limited to those induced by intense winds; however, wind-induced damages are the most prevalent causes of damage during hurricanes, as suggested by a review of the literature (Dunn et al. 2018; Panteli et al. 2017).

The hazard component simulates different hurricane categories and also the historical wind speed of Hurricane Ike and Hurricane Harvey. The wind speed model is obtained from the HAZUS-MH wind model (Vickery et al. 2006). The wind model probabilistically generates the full profile of wind speed during the duration of a hurricane event for various scenarios of hurricane categories. This model generates wind speed values to a scale at the census tract level. The maximum wind speed values for each tract are determined based on their distance to the center of the hurricane.

## 2.3 Power network agent

### 2.3.1 Network structure

The hurricane wind model poses stress on the power network and could cause multiple disruptions to the power network. The power network is a connected grid consisting of elements such as generators, substations, transmission lines, poles, conductors, and circuits. The data for modeling actual power networks within an area are either unavailable or difficult to access due to the security issues. Therefore, the power network in this study is modeled by using a synthetic power network introduced by Birchfield et al. (2017) and Gegner et al. (2016). The implemented synthetic power network represents a near-real representation of the power network in the study area, which matches the topological characteristics of the actual network in Harris County. The synthetic power network determines the location of the synthetized generation and loading substations based on the required loads and the publicly available power plant data. Then, the substations and generators are connected by transmission lines through a network that has structural and topological properties of an actual network and a converged power AC flow.

The distribution network consists of distribution poles and conductors. The number of distribution poles is estimated based on the population of each tract assuming each pole serves 40 customers (Ouyang and Dueñas-Osorio 2014). The poles are directly linked via a distribution line to the distribution pole. Similarly, each distribution pole is connected to households through conductors. This methodology enables investigating damage to the power network in the absence of real data to model the actual system. Components of the power network, including power generators, substations, transmission lines, the distribution network, and their linkages are captured in the modeled synthetic power network.

Failures in the power network occur not only due to the direct damage to the power network component due to wind forces, but to connectivity loss and cascading failures. Figure 2 shows the overview of the failure-modeling process in a power infrastructure network. The model includes two elements capturing the failure of the network from its exposures to a hurricane: 1) *component damage*: failure in the power network components, which is modeled by incorporating fragility functions (also called fragility curves). The fragility functions help determine the probability of damage to the network component based on hazard intensity. 2) *connectivity disruptions*: The failure of a network component may lead to a series of consecutive connectivity losses. We used connectivity analysis of the network to model such cascading failures in the power network.

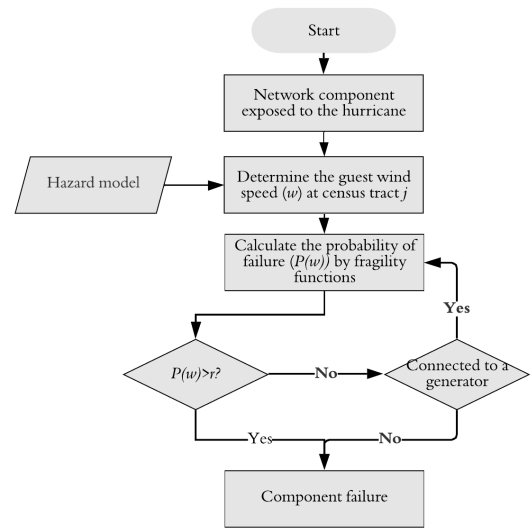

**FIGURE 2.** Schematic overview of the process for modeling the power system failure

### 2.3.2 Component damage

Fragility curves are used to model the failure in the components of the power network. Fragility curves are commonly used for modeling damages to infrastructure systems in response to natural hazards (Winkler et al. 2010). Fragility curves, in this model, determine the failure probability ($P(w)$) based on the imposed wind speed. To this end, the failure probability would be compared to a random variable $r \in [0,1]$ from a uniform distribution, in each iteration. A component, such as power poles, would fail if the failure probability becomes greater than the generated random number. In this model, we consider the failure in the critical components of the power network: substations, transmission lines, distribution poles, and conductors. Damage to power plants by hurricanes, being highly unlikely, was not been considered as structural damage (Ouyang and Dueñas-Osorio 2014).

*Substations*: The damage to substation loads are modeled by implementing the aggregated fragility functions developed in HAZUS-MH 4 (FEMA 2008). The fragility functions provide failure probability based on the local terrain, wind speed at the area, and the structural characteristics of the substation. Equation 1 shows the general form of the fragility function. In this equation, the probability of failure ($P_f$) is related to the exposed wind speed ($x$). The two parameters, mean ($\mu$) and variance ($\sigma^2$) are used to define the lognormal fragility curve. The fragility curves used for modeling damage to the substations are plotted in Figure B-4 in the Appendix B.

$$P_f(damage|w=x) = \int_x^{-\infty} \frac{1}{\sqrt{2\pi}\sigma} exp\left(\frac{-(ln(x)-\mu)^2}{2\sigma^2}\right) d_x \quad (1)$$

*Transmission elements*: Transmission elements include the transmission lines and the transmission towers, which support the lines. The length of the transmission is determined based on the specific latitude and longitude of the generators and substations loads in the synthetic network. The number of necessary transmission towers is estimated by assuming 0.23 km between two consecutive towers. Similar to the fragility function in Equation 1, we implemented a lognormal fragility function for determining the ($P_f$) of the transmission towers. The implemented fragility curves for modeling damage to the transmission tower are shown in Figure B-2 in the Appendix. Damage to transmission towers is modeled so that towers fail independently of one another (Panteli et al. 2017); therefore, the total failure probability for the transmission element due to damage to the support structure between two substations which have *n* towers would be calculated using the following steps. In Equation 2, $P_{T(w)}$ is the probability of failure in the transmission element, $P_k$ represents the probability of failure of an individual tower between substations, and $N$ is the number of required towers for supporting the lines.

$$PT(w) = 1 - \prod_{k=1}^{N}(1-P_{k,w}) \quad (2)$$

Extreme weather conditions could cause great damage to transmission lines; thus, separate fragility curves are used to model such damage. Following the approach by Panteli et al. (2017), a linear fragility function, as shown in Equation 3 and Figure B-2 (Appendix B) is implemented for calculating the probability of failure for the transmission lines.

$$PL(w) = \begin{cases} 0.01, & if\ w < w_{critical} \\ PL, & if\ w_{critical} < w \leq w_{collapse} \\ 1, & if\ w \geq w_{collapse} \end{cases} \quad (3)$$

This equation considers three conditions. First, if the wind speed is below a certain level of "good weather condition," the probability of failure is much lower (0.01). Here, $w_{critical}$ is the wind speed at which the transmission lines start can sustain damage, and $w_{collapse}$ represents a situation when the survival probability of the component is very small. Then the component's probability of failure ($PL$) is calculated by considering a linear relation in the intermediate phase between $w_{critical}$ and $w_{collapse}$. These wind speed thresholds are assumed to be between 30 m/s and 60 m/s following empirical studies (Murray and Bell 2014; Panteli et al. 2017). In the presence of data from utilities, the equations and thresholds could be adjusted to reflect the real behavior of the components; the pseudo algorithms are presented in Tables A1 in the Appendix A.

*Distribution elements*: The synthetic distribution network considers the failure of the conductors that connect the households to the power network and the poles that support the conductors. The empirical damage models, developed by Quanta Technology and implemented by Quanta (2009) and {Formatting Citation} are used in the absence of field data. The fragility equation for modeling the failure to the conductors is shown in Equation 4. This equation (also see Figure B-3) draws the relationship between the wind speed ($w$) and the probability of failure to the conductors ($PC(w)$) in the distribution network.

$$PC(w) = 8 \times 10^{-12} \times w^{5.1731} \quad (4)$$

Lastly, the fragility function for modeling failure in the distribution poles is implemented in the model. Several studies have developed fragility equations for the distribution poles depending on their material, age, and maintenance (Salman et al. 2015; Salman and Li 2016; Shafieezadeh et al. 2014). The fragility equation developed by Shafieezadeh et al. (2014) is used in this study to model the failure in the distribution poles in this study. An example of the fragility curves is shown in Figure B-3 in the Appendix B.

### 2.3.3 Connectivity disruption

The failure of a component in the power network may propagate through the network and lead to connectivity loss (also called cascading failures) (Winkler et al. 2010). The model also considers the cascading failures due to the interdependencies among the components of the power network. For example, when a substation experiences damage, if the distribution network elements connected to the damaged substations are no longer connected to a power generator through other network components, these subsequent distribution networks would also be removed from the power network (Mensah and Dueñas-Osorio 2016). Therefore, at each iteration of the model, the connectivity of the subsequent network component to a generator will be assessed. The pseudo codes of the developed algorithm are shown in Table A-2 in the Appendix A.

### 2.3.4 Restoration process

Restoration activity takes place after the hurricane passes through the affected area. After the failures in the power network are detected, the utility repairs damaged components of the power network. The downtime of different system elements depends on three main factors: (1) the extent of damage to the power network, (2) the available resources to the utility for restoring service, and (3) the utility's strategy for restoring the power (Duffey 2019; Liu et al. 2007). Severe hurricanes pose more danger to the infrastructure elements and make it difficult for the utilities to restore services. The number of crews and spare equipment in place also affect the restoration time (Xu et al. 2019). Finally, the priority of restoration activities influences the duration of outages. For example, restoration in more populated areas may sometimes be prioritized to meet the needs of a higher number of affected households (Liu et al. 2007).

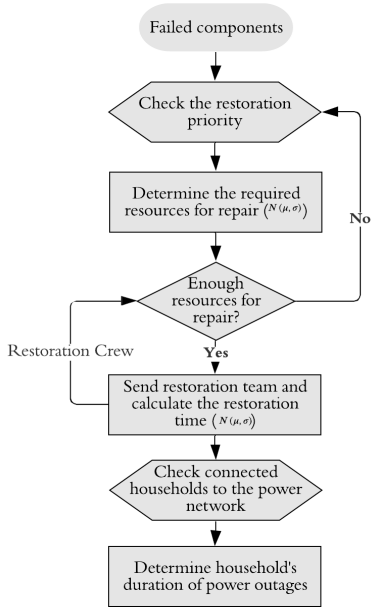

**FIGURE 3.** Schematic overview of the process for modeling the restoration activity

To determine restoration duration, the model determines the duration of the power outages by considering the dynamic repair process (Figure 3). The process involves multiple steps (Sharma et al. 2020). First, the priorities are given to the power restoration in different areas to implement repair and restoration strategies. Then, for each damaged element, the required resources and time to repair will be calculated based on Table B-1 (Appendix B). The time to restore each element is calculated by considering a normal distribution, $N(\mu, \sigma)$, with specific mean and standard deviation (Mensah 2015). The resources in this model are crews, material, and machines. The number of teams needed for the repair task is given in Table B-1. The utility could have a finite number of resources in place, but then these resources could be augmented daily by assistance from other utilities through Regional Mutual Assistance Groups (RMAGs) and collaborations (Edison Electric Institute 2016). A linear relationship is assumed for the increase in repair resources (Figure B-1) based on the results of previous resource modeling studies (Ouyang and Dueñas-Osorio 2014). The model inputs resources and initially implements 800 teams increasing by 15 teams per hour for a week as the base case scenario.

### 2.3.4 Restoration strategies

Based on a review of the literature, there is no standard way of restoring power when a severe weather event damages a power network (Applied Technology Council 2016). Some utilities would prioritize the restoration of the service areas with greater populations; however, this restoration strategy might favor residents living in a larger metropolitan area and might adversely affect people in rural areas (Liu et al. 2007). Other strategies mainly focus on physical characteristics, such as prioritizing the components with a high (Liu et al. 2021; Ouyang and Dueñas-Osorio 2014). The model uses priorities assigned to the components in the network to generate the different repair strategies.

In this study, we tested the influence of three main strategies for restoring the power for residents, *component-based restoration*, *population-based restoration*, and *social vulnerability-based restoration*. In component-based restoration, the model prioritizes the restoration of critical components, such as failed substations and transmissions. After repair of these components, the model initiates the repair of the damaged distribution network comprising conductors and poles in a random manner. This activity represents prioritization of the repair of the component, which serves areas with higher population or higher social vulnerability scores informed by census data and a socially vulnerability index (SVI) (Flanagan et al. 2011). Selecting these strategies enables ranges of service restoration duration in different areas. Therefore, in this model, households would experience varying levels of power outages due to the differences in the extent of damage, their tolerance for disruptions, and utility's restoration strategy depending upon neighborhood.

### 2.4 Household agent

Households have varying levels of tolerance for withstanding power outages. Empirical data from household surveys from three major hurricane events (Harvey, Florence, and Michael) were used in conjunction with theoretical decision-making models to simulate the underlying mechanisms affecting the tolerance of households. The tolerance depends on households' decisions about the protective actions and their inherent needs for the service (Baker 2011; Coleman et al. 2020; and Esmalian et al. 2020b). The model includes the process through which households know about the event and form perceptions about the risks. Then empirical models developed based on the survey data used in concert with decision-making processes are used to determine the probability of a household taking protective actions. This probabilistic characteristic of the households' behaviors enables consideration of the uncertainties regarding the individual's behavior in the model. Finally, the household's hardship status would be determined based on tolerance and the duration of outages. The pseudo algorithms are shown in Table A-3 in the Appendix A.

### 2.4.1 Information propagation process

Two information propagation processes are considered in this model (Figure 4). First, we modeled information sharing through the official sources (such as mass media). In the days before hurricane landfall, officials disseminated information about the upcoming hurricane, which is modeled by implementing a probability of receiving the information by the households ($P_o$). In addition, those who receive the message might also share the information with their immediate social network, depending on how important they perceive the risks of the hazard, and then take protective action themselves. Hence, two elements of ($P_i$) and ($P_n$) are considered for implementing the information-sharing process by households. Those who perceive a great risk from the hazard and take protective actions are more likely to share information with their social network than those who do not perceive a high risk. These probabilities are determined using the empirical data and considering a higher value for the probability of receiving information from the officials, and then information sharing from those who decide to take the protective actions, and finally the probability of sharing information through those who do not get prepared.

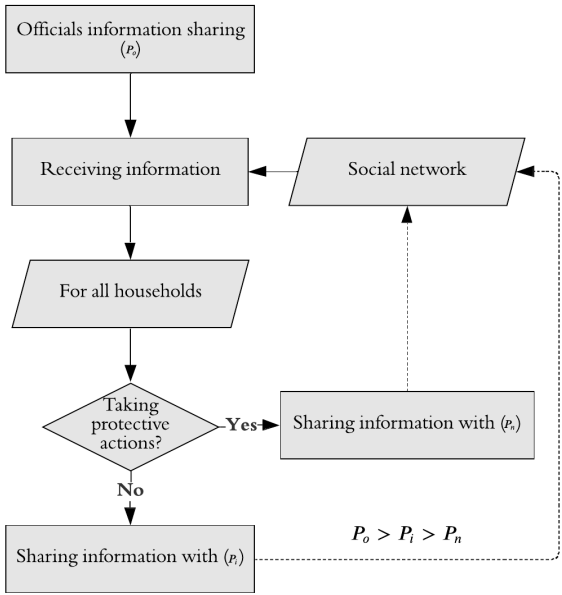

**FIGURE 4.** Schematic overview of the information seek/share behavior

### 2.4.2 Household agent's social network

Agents interact with each other and influence the decisions of others through their social networks. The social network of the agent would not only influence the information propagation process; it would also affect other agents' decisions regarding the protective actions (Anderson et al. 2014; Tran 2012). Multiple network structures—random network, small-world network, scale-free network, and distance-based network—characterize how households are connected with each other. These network structures are present in real-life social settings.

For example, the literature suggests that information sharing through online social media, which follows a scale-free network structure, could expedite information propagation (Nocaj et al. 2015; Schnettler 2009). Therefore, we considered multiple network structures to account for various modes through which households could interact and share information, and we tested the impact of such structures on the overall impact of the hazard on the communities. The social network would affect both the information propagation process and the household decision-making on the protective actions through peer effect.

### 2.4.3 Household agent's risk perceptions

Household agents form a perception about the potential duration of the power outages. We analyzed data collected from the household surveys to determine households' expectations of the disruptions; the summary statistics of household survey data could be found in (Esmalian et al. 2020b). Households' expectation of the duration of the disruption affects their decisions regarding taking protective actions. Those with higher expectations of the disruptions are more likely to take protective actions (Coleman et al. 2020; Lindell and Hwang 2008). Equation 5 shows how the mean of the duration of the expectation is related to the predictors through a *log* link by implementing a Poisson regression model. In this model, $x_f$ refers to the forewarning duration of the event, $x_i$ shows if the households receive the information about the hurricane, $x_o$ is home ownership, $x_a$ captures whether the head of the household is elderly, $x_m$ captures if the household member has a mobility issue, and $x_{fz}$ refers to if the households live in a flood zone.

$$\mu = exp[1.74700 + 0.30471 log(x_f + 1) + 0.12369 x_i - 0.27720 x_o - 0.21065 x_a - 0.51210 x_m - 0.28153 x_{fz}] \quad (5)$$

### 2.4.4 Household agent's socio-demographic characteristics

Households' demographic characteristics influence their perceptions of the risk, decisions regarding the protective actions, and consequently, their tolerance for the disruptions (Baker 2011; Coleman et al. 2019; Horney 2008). In this model, household's demographic characteristics are considered by developing a sample of agents based on publicly available census data. A population is sampled by considering the probability of being from a specific segment of a community by using the actual proportions in the census data. In particular, data income level, race, age, educations, mobility/disability conditions, and type of housing of the households were collected. In addition, to determine whether a household was in a flood zone, household location was plotted against a 500-year-flood map.

The demographic characteristics of households not only influence their decisions on protective action, but also affect households' level of need for the service. The level of need is modeled through the use of empirical data. In the surveys, this variable is measured with an ordered five-level Likert scale; therefore, a cumulative logit model is developed for determining the level of need (Equation 6). The model relates the effect of predictor $x$ on the log odds of response category $j$ or below by coefficient $\beta$ (Agresti 2007). This type of modeling helps in determining the probability of $Y$ (the level of need) falling below a certain level (Equation 7). Then, as the probability of all levels sums up to 1, the probability of each level could be determined. Appendix B outlines the models for estimating the level of needs.

$$logit\, P(Y \leq j) = \alpha_j + \beta x \qquad (6)$$

$$logit[P(Y \leq j)] = log\left[\frac{P(Y \leq j)}{1-P(Y \leq j)}\right] = log\left[\frac{\pi_1 + \cdots + \pi_j}{\pi_{j+1} + \cdots + \pi_J}\right], \quad j = 1, \ldots, J-1 \qquad (7)$$

### 2.4.5 Household agent's protective action process

Households take protective actions for reducing the impacts of power outages in two ways. First, the general preparedness behavior of households in terms of obtaining food, water, and emergency kit supplies helps them to better cope with the outages. Second, some households might take further actions by purchasing a generator. We modeled the protective action process of households by implementing the diffusion model developed by Banerjee et al. (2013). As shown in Figure 5, households are first informed about the hurricane through the information propagation of officials or their immediate social network. Second, an initial number of households decide to take protective actions depending on their decisions' probability ($P_p$). Households' probability of taking protective actions ($P_p$) depends on the households' personal characteristics, such as demographic characteristics, risk perception, and peer influence. Equation 8 shows the implemented logistic function to model this process. Third, those who decide to take protective actions influence their social network by passing the information regarding their protective actions. Forth, the newly informed households now decide if they want to take protective action. This process initiates as soon as the officials detect the hurricane and ends after ($f$) days of forewarning.

$$log\left(\frac{P_p}{1-(P_p)}\right) = X_i \times \beta + \lambda \times F_i \qquad (8)$$

In this model, $\beta$ is the vector of the coefficients that relate the personal characteristics ($X_i$) to the log-odds ratio of the protective action decisions. $F_i$ is the fraction of the household's social network who had decided to take protective actions divided by the total number of household's social network. $\lambda$ represents the change in the log-odds ratio of protective actions due to peer influence. A value of zero for $\lambda$ describes the case in which households make their decision independent of their social network, while larger values of $\lambda$ refers to a situation when households affect the decision of their social network. The empirical models were implemented to determine the $\beta$, and the model has been tested to determine the range of $\lambda s$. Details related to the factors considered for developing these models are presented in Appendix B.

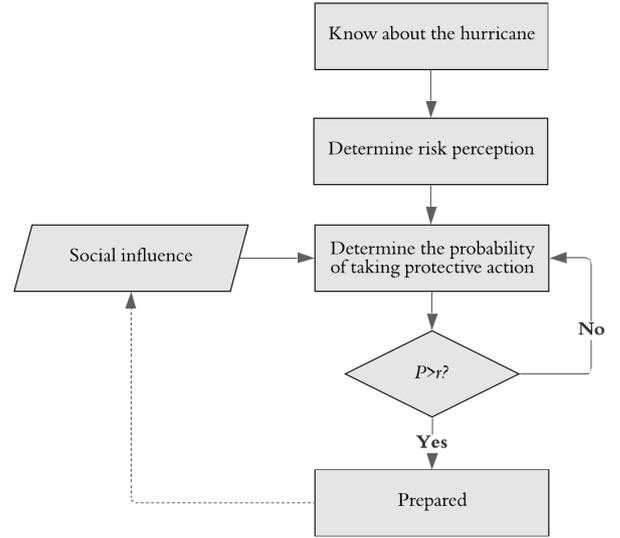

**FIGURE 5.** Schematic overview of taking protective actions by households

### 2.4.6 Household agent's protective action process

Households have different levels of tolerance for withstanding the prolonged power outages (Esmalian et al. 2019). This is why even a similar outage duration would cause varying levels of hardships on different households (Coleman et al. 2019). Households' tolerance for power outages is a function of their protective actions and inherent needs for the service. Household tolerance is determined by implementing accelerated failure time (AFT) models, which are a type of survival analysis approaches for the time-to-event data (Dale 1985). This type of modeling was found to best describe the model and to have the lowest prediction error when compared to generalized linear models (Poisson family and negative binomial regression) and ensemble learning methods (random forests and boosting) (Esmalian et al. 2020a). Using AFT models, we can directly relate tolerance to the predictors with a linear relationship, as shown in Equation (9).

$$log\, \mu_i = x_i^T \beta + \varepsilon_i \qquad (9)$$

where, $\mu_i$ represents the mean tolerance, $x_i^T$ denotes the vector of predictor, $\beta$ is the vector of parameters, and $\varepsilon_i$ is an error term that is assumed to be independently distributed. In this model, three main predictors were used for determining tolerance: households' level of need for the service, their preparedness for the event, and if they obtain a generator to withstand the power outages. The protective actions of the households are determined through a probabilistic approach outlined in the previous sub-section. The level of need is determined based on their socio-demographic characteristics to be considered in calculating the tolerance level.

In the last step, the households' experienced hardship is determined by integrating the results from the restoration process with households' tolerance. Households experience different levels of duration of disruptions and experience hardship when the duration of the outage exceeds their tolerance. Figure 6 presents the process for determining whether the households' experience hardship from the service disruptions.

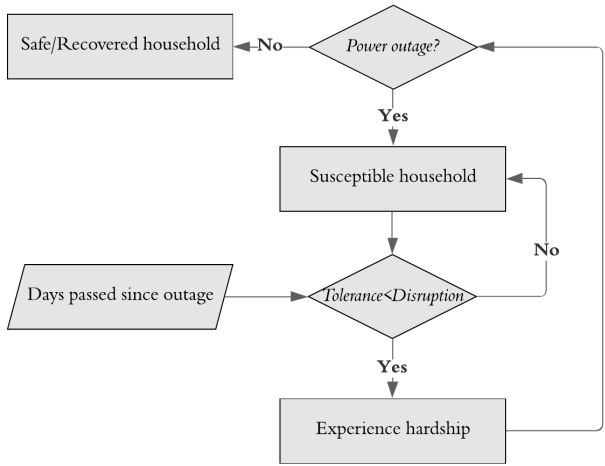

**FIGURE 6.** Schematic overview of household hardship experience process

## 3 MODEL IMPLEMENTATION AND SIMULATION EXPERIMENTS

### 3.1 Computational Implementation

Computational representation of the proposed multi-agent modeling framework includes developing and implementing algorithms and mathematical models to capture the theoretical logic representing the experienced hardship of households due to disaster-induced disruptions. The computational model is created by using an object-oriented programming platform, AnyLogic 8.3.3. Figure 7 depicts the Unified Modeling Language (UML) diagram of the model, which shows the class of the agents, agents' attributes and functions, and their relationships. A sample of 2500 households based on the demographic characteristics of Harris County were generated and placed in the census tracts. The synthetic power network includes a total of 97 substations, 242 transmission elements, and 1433 distribution elements located in Harris County based on latitude and longitude coordinates as described in the power network agent section.

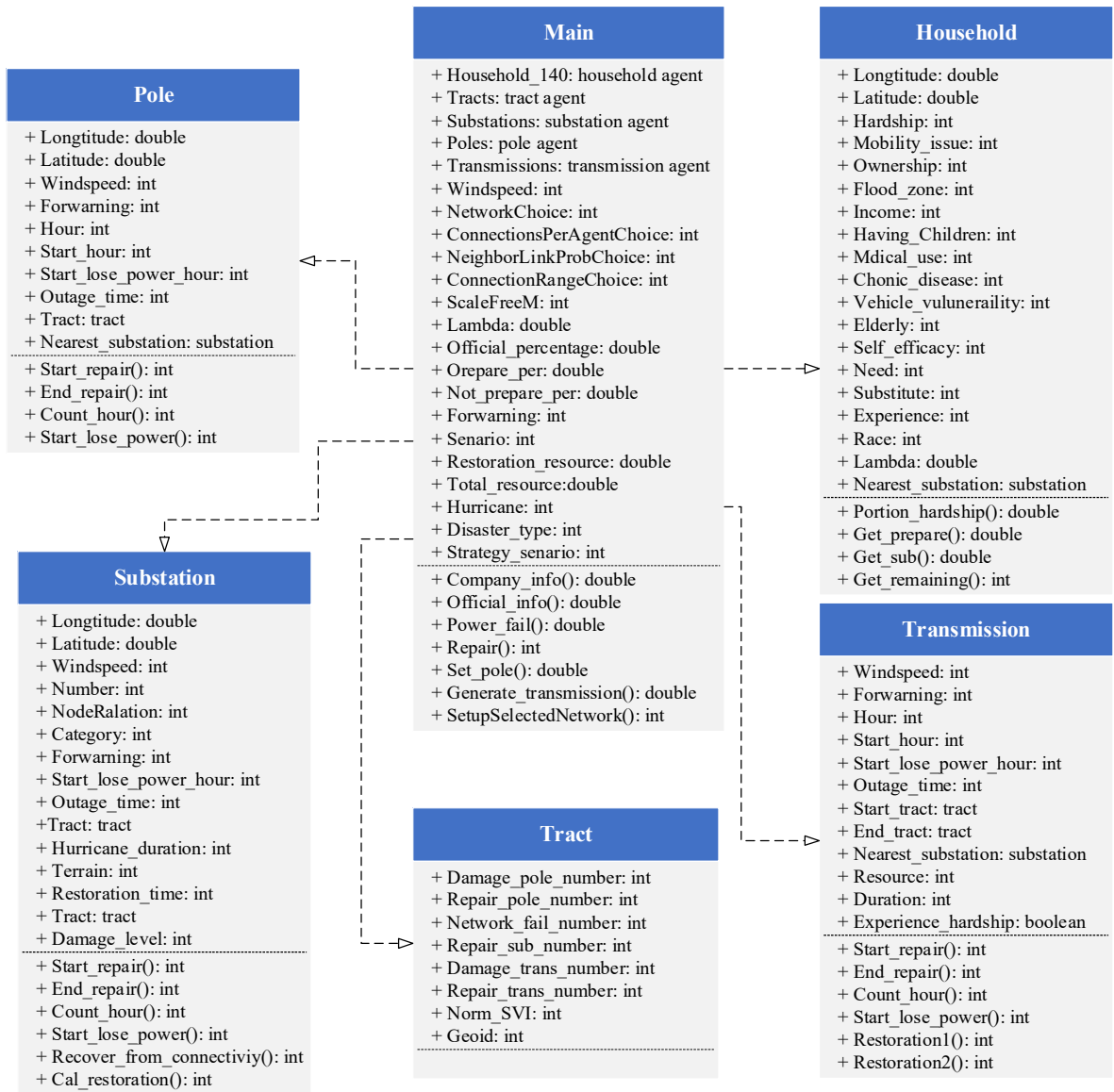

**FIGURE 7.** UML class diagram of the multi-agent simulation model

## 3.2 Verification and Validation

The model is verified and validated through a systematic and iterative process to ensure the quality and credibility of findings. Various internal and external approaches were conducted to verify the data, logic, and computational; algorithms in the simulation model (Bankes and Gillogly 1994; Mostafavi et al. 2016; Rasoulkhani et al. 2020). First, the internal verification of the model was ensured by using the best available theories and standard approaches for implementing the models' logic and rules. Second, we used reliable empirical data collected in the aftermath of three major hurricane events to develop the model. Furthermore, we conducted a component validity assessment for ensuring the model components' completeness, coherence, consistency, and correctness. The extreme conditions were tested to examine the model's ability to generate reasonable outcomes. External verification of the model was ensured by examining the causal relationships among the model components. The behavior of these sub-components under different values was traced to ensure the external verification of the model. The model logic and functions were examined to discover any unusual patterns to ensure that logic and assumptions in the model are correct.

Furthermore, results from similar studies and reports on the impact of hurricanes on the power networks were used to validate the model's output for the physical system (Mensah and Dueñas-Osorio 2016; Ouyang and Dueñas-Osorio 2014). For validation, the generated patterns in the model outputs were compared against the empirical data to validate model behavior. The mode of each simulated output was used to determine the system's behavior, then the generated patterns from the model were compared with the actual household behaviors from the survey empirical data and similar studies and reports. In this study, the intent of the model was to examine the strategies to reduce societal impacts of power outages. In particular, emergent behavior patterns of the outputs were of interest. The model is capable of generating patterns and values similar to the empirical data (Figure 8). The model outputs capture the Hurricane Harvey scenario in Harris County, Texas, in 2017 (Figure 8). For example, the generated proportions of households which prepare and obtain substitute energy sources (generators) are similar to those values from empirical data. Some differences arise in the model results for large and small values of the forewarning time; however, the distribution of tolerance is close to the empirical values. It is worth mentioning that the primary objective for creation and use of multi-agent and agent-based models is not prediction, but rather to generate examples of the probabilities of various possibilities for robust decision making under uncertainty (Mostafavi et al. 2016).

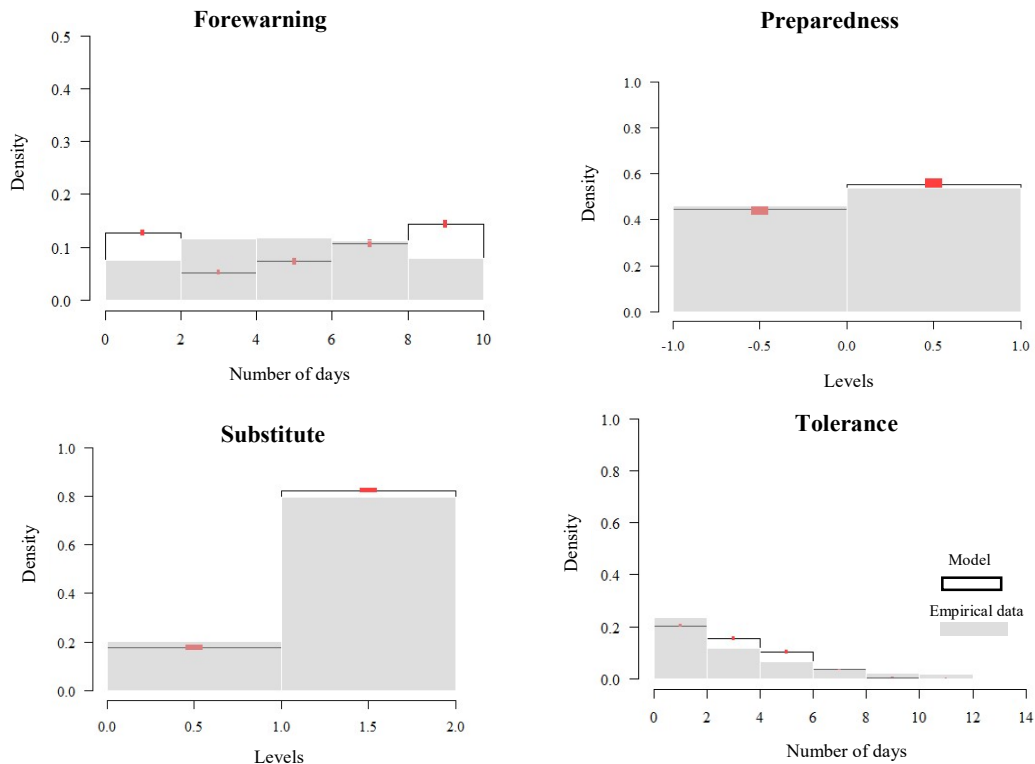

**FIGURE 8.** Comparing values generated through the model with the empirical data. The red whiskers show the 5% and 95 % values of the model replications.

## 3.3 Model Output Description

The percentage of households experiencing hardship from power outages is recognized as an indicator of the societal impacts on the community. When a households' duration of power disruption exceeds their tolerance, they experience hardship. This indicator includes both the physical impact and the societal susceptibility of the households for the risk posed. This dynamic measure is calculated for all households based on their location and their tolerance during the time without service. Figure 9(a) shows how the dynamic profile of hardship could be implemented to assess the effectiveness of various strategies in reducing the societal impacts of power disruptions. Different scenarios could be tested to find ways to mitigate the societal risks of disruptions to power networks.

In addition to examining the societal impact on the community, the model enables examining the impact on various sub-populations (Figure 9(b)). This capability of the model enables understanding of whether system restoration strategies are equitable. For example, while one strategy might reduce the societal impact on the community as a whole, it is possible that the strategy is in favor of certain demographics in the community. Thus, strategies would be examined to determine how they improve the condition of different social groups in the affected community.

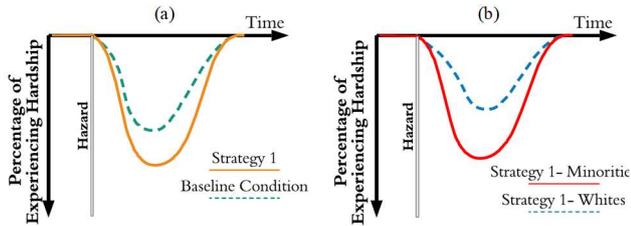

**FIGURE 9.** Schematic dynamic profile of hardship. (a) comparison of the effect of strategies, (b) comparison of the impact of strategies on different social groups.

## 3.4 Simulation experimentation

The developed simulation model enables testing scenarios through various variables such as household characteristics, household social network structure, forewarning duration, hurricane category, and restoration units and strategy. The user could choose the values related to each of these variables in an interactive user interface (Figure 10a). The model outputs the various values related to different variables, including household protective actions and tolerance, extent of damages to the different components of the power network, and the households' profile of hardship. In addition, as shown in Figure 10b, the model visualizes the spatial distribution of households' states by color-coding them depending on their states. Households who experience the power outages are shown in orange, those whose tolerance becomes less than their duration of disruption and experience hardship are shown in red, and the color changes to green when the power is restored for these households.

We performed Monte Carlo experimentation in the scenario testing to account for the stochasticity in the model. The primary variable of interest in the model experimentation was the percentage of the households who experience hardship from the power disruption. Therefore, experiments were replicated as many times as the mean value of proportional of households experiencing hardship reached 95% confidence interval with 5% error (Hahn 1972). The experiment scenarios were designed by changing the input values of each scenario and replicating iterations for each of the experiments.

## 3.5 Simulation experimentation

The model is implemented for scenario testing aiming at (1) identifying the combination of the strategies which would lead to the mildest societal hardship due to the power outages, and (2) examining the extent to which the strategies are equitable. In this study, we examine three main strategies to reduce the societal impacts of power outages. First, the power utility's restoration strategy would be evaluated to examine its influence on the hardship levels. In this regard, three strategies of restoration based on the importance of the components, population size, and social vulnerability index would be evaluated. SVI is a widely adopted measure for examining the susceptibility of populations in disasters. Second, the influence of the forewarning time on the experienced hardship of the households would be examined. This evaluation would determine the value of identifying an impending hurricane and communicating critical information with the population. Early warning about the upcoming hazard influences the protective decisions of households (Cremen and Galasso 2021; Watts et al. 2019). Third, the impact of the social network of the households on their experienced hardship would be evaluated. This assessment would show the value of using alternative social networks (such as social media) for disseminating hazard information. Social media platforms, for example, have distinct network characteristics which enable quicker information sharing without spatial boundaries (Watts et al. 2019; Zhang et al. 2019). Therefore, the type, density, and weight of the social influences would be examined to explore their effect on reducing the impacts of power outages on the households. The combination of these strategies to lower the hardship experienced by the households was also examined.

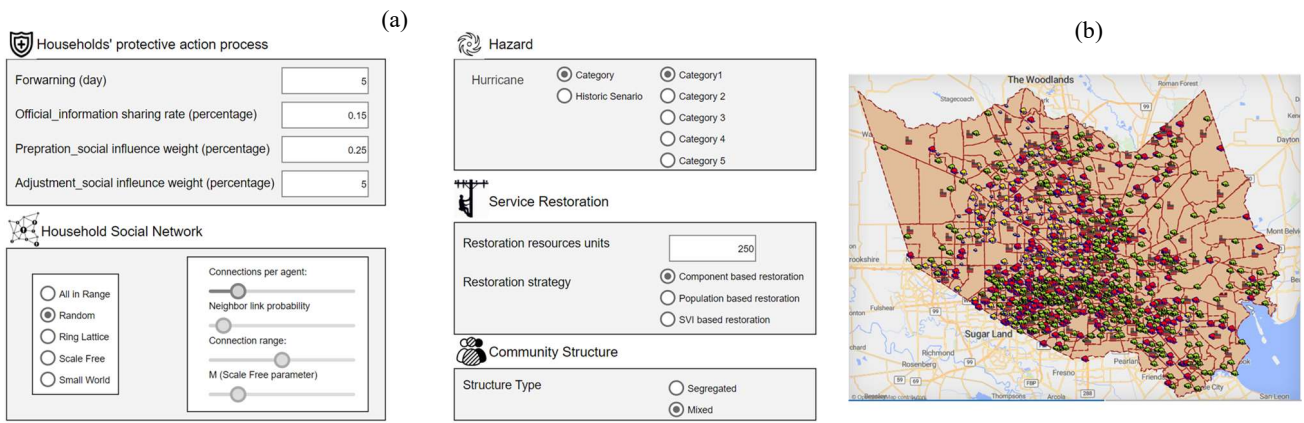

**FIGURE 10.** Screenshots of the developed simulation model.

## 4 RESULTS AND DISCUSSION

The hardship experience of households from scenario analysis was used for exploratory analysis of societal risks of prolonged power outages. The analysis included: 1) examining strategies for reducing the societal impacts and examining to what extent they are equitable, including restoration strategies, forewarning, and social networks; 3) robustness of the strategies for reducing the societal impacts under different scenarios; 4) identifying pathways that lead to low societal impacts. To this end, in a base scenario similar to Hurricane Harvey context with a forewarning of 9 days, the utility implements the component-based restoration, forewarning of 9 days, in a scale-free social network between households. Scenarios were then modeled and compared with the base-case scenario through Monte Carlo simulation. In the simulation results, day zero is the time when an impending hurricane is identified by the officials as a threat and the information is communicated with the residents.

### 4.1 Simulating community-scale societal impacts

A baseline scenario of societal impact of power outage disruption in a community similar to Harris County affected by a category 4 hurricane is shown in Figure 11. Figure 11(a) shows the mean proportion of households experiencing hardship each day. The results suggest that at maximum, around 50% of the community experienced hardship from the outages, and it took roughly 20 days for the community to fully recover (recovery is determined by having power restored for all households). The impact, however, was not equal among the subgroups in the community. Racial minority groups experienced a higher hardship from disruptions. Figure 11(b) shows the overall probability of experience hardship for each group. This result suggests that racial minority groups are more likely to experience hardship from power outages in comparison with others in the base case scenario. The results overall show the model's capability to capture the societal impact of the disruptions on communities and also reveal the inequities in the impacts of prolonged power outages on vulnerable populations (e.g., minority groups).

### 4.2 Examining strategies for reducing societal impacts

#### 4.2.1 Restoration strategy

Results for comparing different strategies for restoring the power (Figure 12) show that while under the component-based strategy, the maximum proportion of hardship in a day is around 54%. This value would be decreased to around 47% under the population- and SVI-based restoration strategies. The results show that overall, a community similar to Harris County, Texas, would benefit from prioritization of the areas with a higher vulnerable population. In this case, the probability of experiencing hardship for the nonvulnerable population increases; however, the reduction in the probability of experiencing hardship for the socially vulnerable groups leads to an overall reduction in the societal impacts. In addition, giving priority to the areas with a higher population results in the reduction of societal impacts on the affected community overall. These findings suggest that overall, the prioritization of areas with a higher social vulnerability level and also with a higher population could lead to the reduction of overall societal impacts.

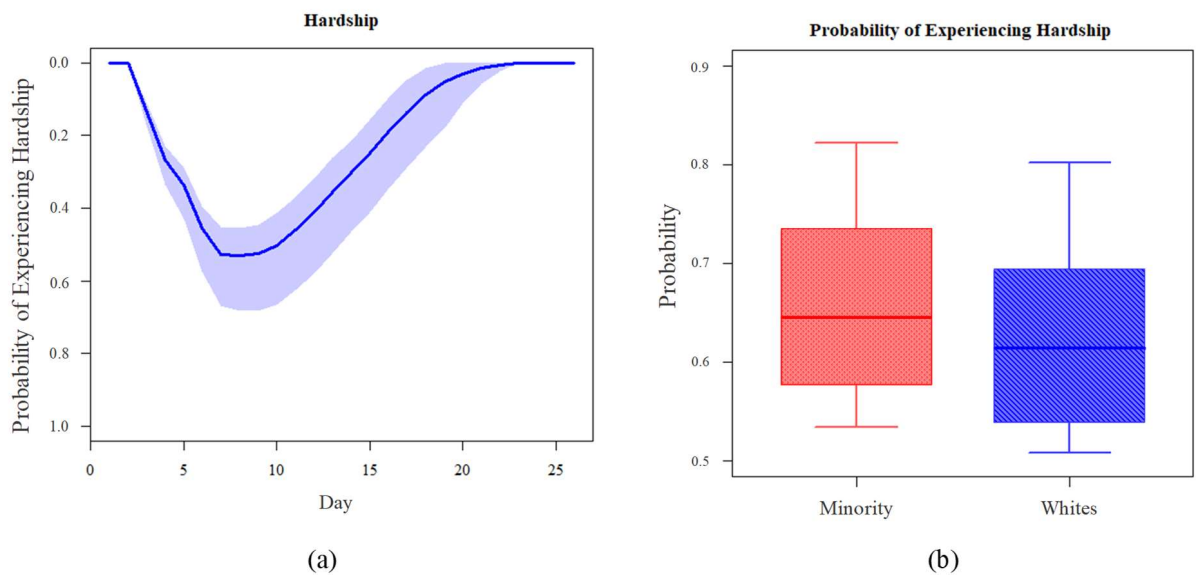

**FIGURE 11.** Societal impacts of disruptions from power outages in the baseline scenario. (a) average daily proportion of households experiencing hardship and the 10% confidence intervals; (b) boxplots and mean lines for the probability of racial minorities and whites experiencing hardship.

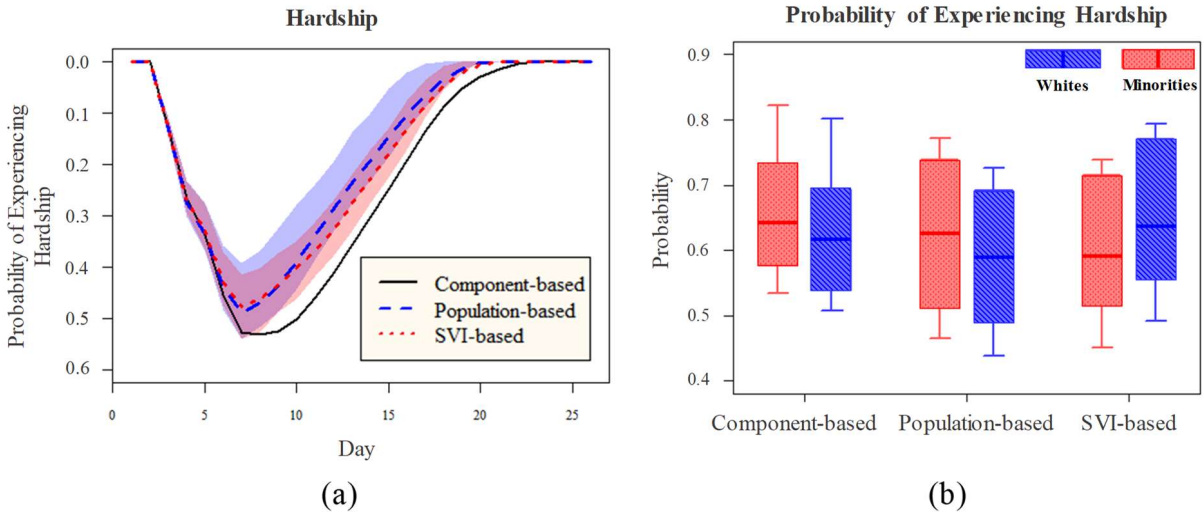

**FIGURE 12.** Comparing different power restoration strategies. (a) dynamic patterns of the proportion of households experiencing hardship under each strategy, with shaded areas indicating the 0.25 and 0.75 percentile of the values; (b) probability of experiencing hardship for different racial groups under each restorations strategy.

The results comparing the effect of different prioritization strategies on racial groups are shown in Figure 13. The charts juxtapose the probability of experiencing hardship for two social groups under different restoration strategies. In the SVI-based recovery, the probability of experiencing hardship decreases by 8% for the socially vulnerable groups while it would increase by 4% for the nonverbal group. The population-based recovery, however, decreases the probability of experiencing hardship by 2% and 4% for the vulnerable and the nonvulnerable group, respectively. The results suggest that the population-based restorations, while improving the overall societal risks, do not favor minority groups. On the other hand, the SVI-based recovery, while increasing the risks for the whites, is enhanced overall restoration strategy. While the population-based restoration and SVI-based would better enhance overall the societal impacts, an SVI-based approach seems to be more equitable.

### 4.2.2 The effect of increasing the forewarning period

Providing a longer forewarning to the communities reduces the societal impacts of power outages to the communities. As expected, the longer duration of the forewarning helps the households to better prepare for the impacts of the power outages and take protective actions to reduce the impacts of power outages on their well-being. Comparing an event with a week of forewarning with a scenario in which the household had two weeks of forewarning, the results suggest that this early identification of a hazard is very effective for reducing the impacts for the communities (Figure 14). The maximum proportion of households experiencing hardship in a day would decrease around 8% when increasing the forewarning time from 7 days to 14 days. Investments in making advancements in predicting and tracking the hurricane pass and proper communication with the household could significantly reduce the societal impacts from the power outages. However, the enhancements in providing longer forewarning would not necessarily reduce the societal impact for socially vulnerable populations. In both the base scenario and the enhanced strategy, minorities show a higher probability of experiencing hardship Figure 14 (b). While the enhanced strategy shows to reduce the impact for the minority groups slightly more than other groups, this strategy seems to treat everyone equally and do not necessarily be in favor of improving the equity in the impact.

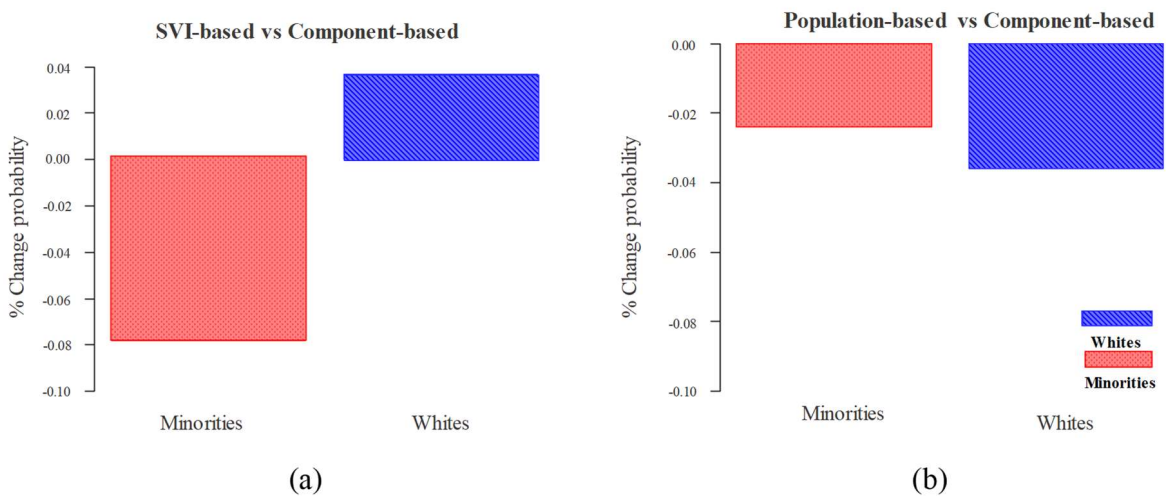

**FIGURE 13.** Comparing the probability of experiencing hardship for the racial groups under each restoration strategy.

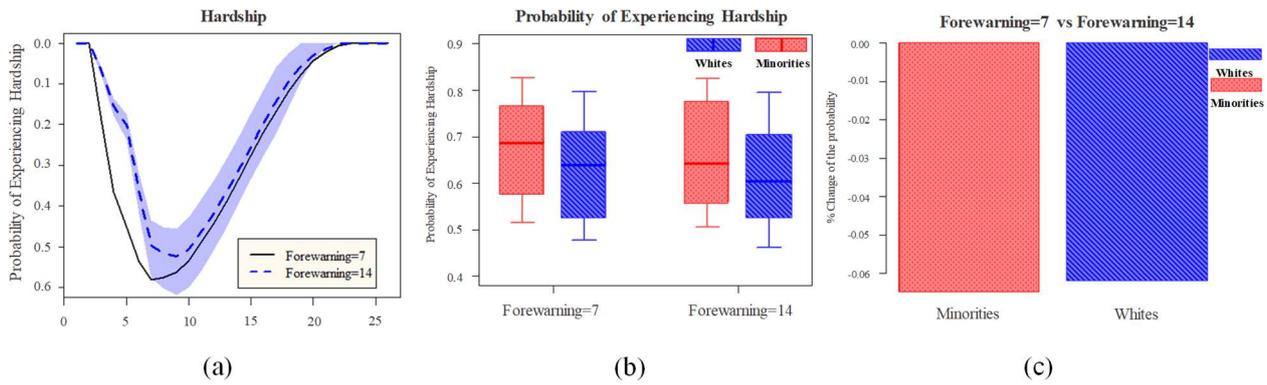

**FIGURE 14.** Comparing different forewarning levels. Figure (a) shows the dynamic patterns of the proportion of households experiencing hardship under each forewarning level. The shaded areas show the 0.25 and 0.75 percentile of the values. Figure (b) shows the probability of experiencing hardship for different racial groups under each forewarning level. (c) shows the change in the probability of expiring hardship for the racial groups under improvement of the forewarning level.

### 4.2.3 The effect of hazard information dissemination and social network types

The social network type has implications regarding which social network people receive information. The two structures of social networks, namely, scale-free and small-world, are compared as each provides certain characteristics in the propagation of information through the community. For example, as discussed earlier, communication among close friends happening offline (in person or on the phone) is through a small-world network, and communication on social media is through a scale-free network (Nocaj et al. 2015; Schnettler 2009). The results from Figure 15 show that there is a slight difference in the societal impacts of power outages on the community when comparing the two network structures. One reason is due to the delays in acting upon the information received by the social network for taking protective actions. Results suggest that the probability of experiencing hardship is greater in the small-work structure. The change in the network structure from scale-free to small-world seems to have a greater impact on the non-minority group. This means that lack of information communication through social media would have more impacts on minority groups compared to white households.

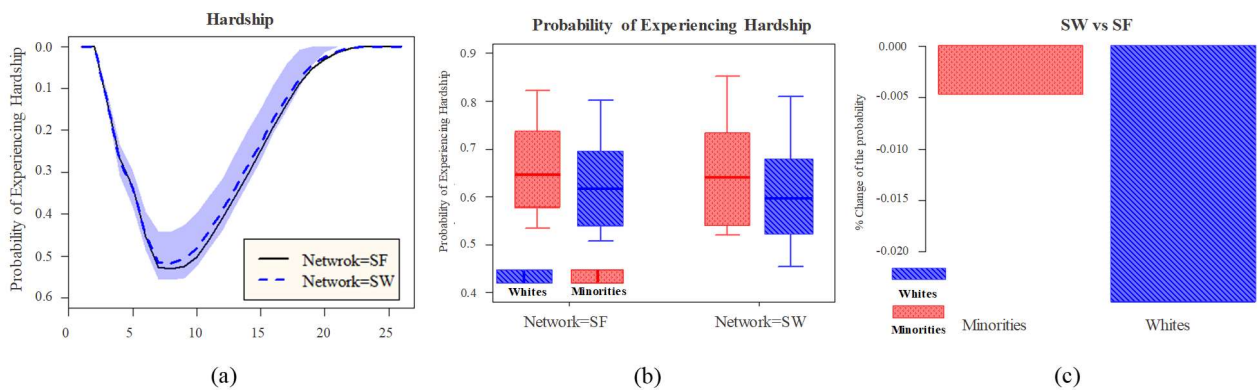

**FIGURE 15.** Comparing scale-free and small-world social networks. Figure (a) shows the dynamic patterns of the proportion of households experiencing hardship under each forewarning level. The shaded areas show the 0.25 and 0.75 percentile of the values. Figure (b) shows the probability of experiencing hardship for different racial groups under each network structure. (c) shows the change in the probability of expiring hardship for the racial groups under a change in the social network structure.

## 4.3 Combined effect of strategies for reducing the societal impacts

## 5 4.3.1 Robustness of restoration strategy to different hurricane categories

The effectiveness of implementing different strategies for restoring power to reduce the societal impacts varies depending on the intensity of the hurricanes. Figure 16(a) and 15(b) show the probability of expiring hardship for each strategy and the dynamic impact under the four hurricane categories. While there is no significant advantage for implementing population-based and SVI-based strategies during low-impact events such as hurricane category 1, these strategies seem to over-perform the component-based restoration during hurricane category 2 and 3. The largest difference is related to hurricane category 3, with population-based restoration leading to the mildest societal hardship. However, the difference between the societal impacts of implementing the population-based and SVI-based with component-based decreases in hurricane category 4. This result suggests that the effectiveness of the improved restoration strategy may not increase linearly as the intensity increases. When the intensity increases to hurricane category 4, SVI-based strategy seems to perform slightly better than population-based and component-based restoration slightly. This trend is due to the increased gap between the vulnerable population and others when the intensity increases as the intensity of the hurricane increases. Figure 16(c) and 16(d) compare the probability of experiencing hardship for the racial groups for population-based and SVI-based relative to component-based, respectively. While the population-based recovery seems to improve the condition for both social groups, this strategy seems to be slightly in favor of the non-vulnerable population slightly. However, the SVI-based restoration reduces the societal impacts for the vulnerable population more than other. Therefore, when the intensity increases to hurricane category 4, this strategy reduces the overall hardship even slightly better than the population-based restorations.

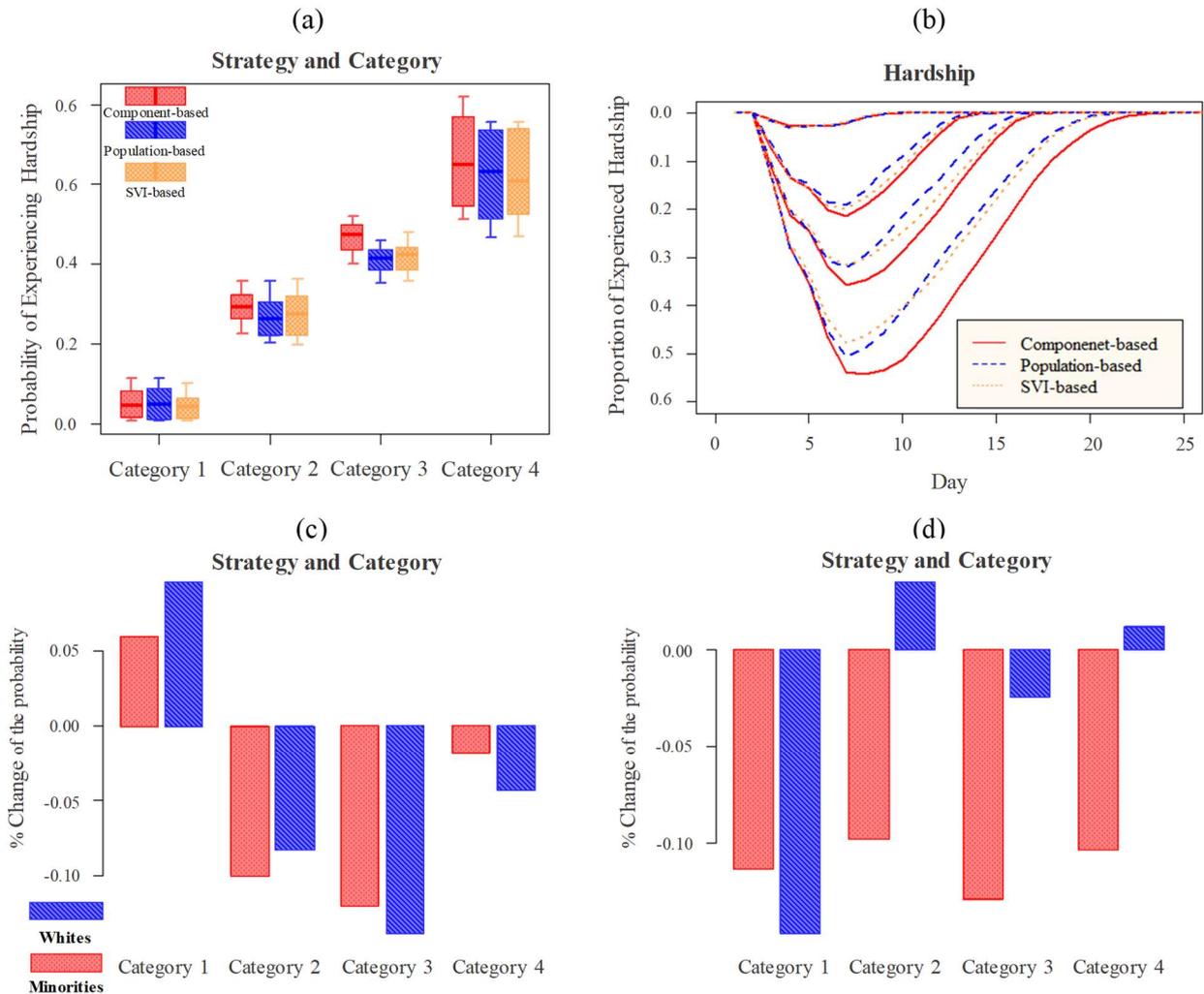

**FIGURE 16.** Effect of restoration strategy on the societal impacts of power outages under various hurricane intensities. (a) shows the histograms of the probability of expiring hardship for each scenario. (b) displays the average daily experienced hardship for each scenario. (c) and (d) show the percentage difference of the probability of experiencing hardship for the racial groups under each scenario.

## 6 4.3.2 Robustness of forewarning to different hurricane categories

The extent of reduction in the societal impacts of power outage by providing a longer forewarning time varies depending on the hurricane category. The reduction of societal impacts provided significantly changes for the forewarning time of more than 6 days (Figure 17(a) and 17(b)). These figures show that both the probability of experiencing hardship and the daily experienced hardship sharply decline when forewarning time increases to more than 6 days. The results explain the major impact of rapid onset hazard events (such as fast-moving hurricanes) on the affected communities. Figure 17(c) compares the probability of experiencing hardship for scenarios increasing by 3-day increments of forewarning. This result suggests that providing longer forewarning is mainly an effective strategy for low-intensity hurricanes. Providing a longer forewarning category 3 and 4 hurricanes seems to diminish. Thus, implementing this forewarning strategy may not solely reduce the societal impacts of high-intensity hazard events. Lastly, Figure 17(d) shows the percentage of reduction of the probability of experiencing hardship for racial groups if the forewarning increase from 6 days to 12 days. The result shows that increasing the forewarning duration does not seem to benefit certain groups. While minorities experience a decrease in the experienced hardship under hurricane categories 1 and 2, the difference does not seem to be significant, especially for the more intense hurricane events.

## 6.1 Pathways to different levels of societal impacts

A combination of scenarios was used to create the scenario landscape (Figure 18) and to evaluate the combination of strategies that lead to the least onerous societal impacts of power outages. To this end, classification and regression tree (CART) analysis was implemented to examine the effect of different variables for reducing the societal impacts under various scenarios (Breiman et al. 1984). CART is a tree-based classification technique that explains how a target variable could be determined based on the interaction among a large number of predictors. This algorithm recursively partitions into binary splits, which maximizes the homogeneity of the groups in relation to the dependent variable (Prasad et al. 2006). The higher splits show the variables with a stronger influence over changes in the dependent variable, which is the experienced hardship in the scenario landscape. CART analysis is shown to be effective in meta-modeling analysis based on simulation results (Mostafavi 2018).

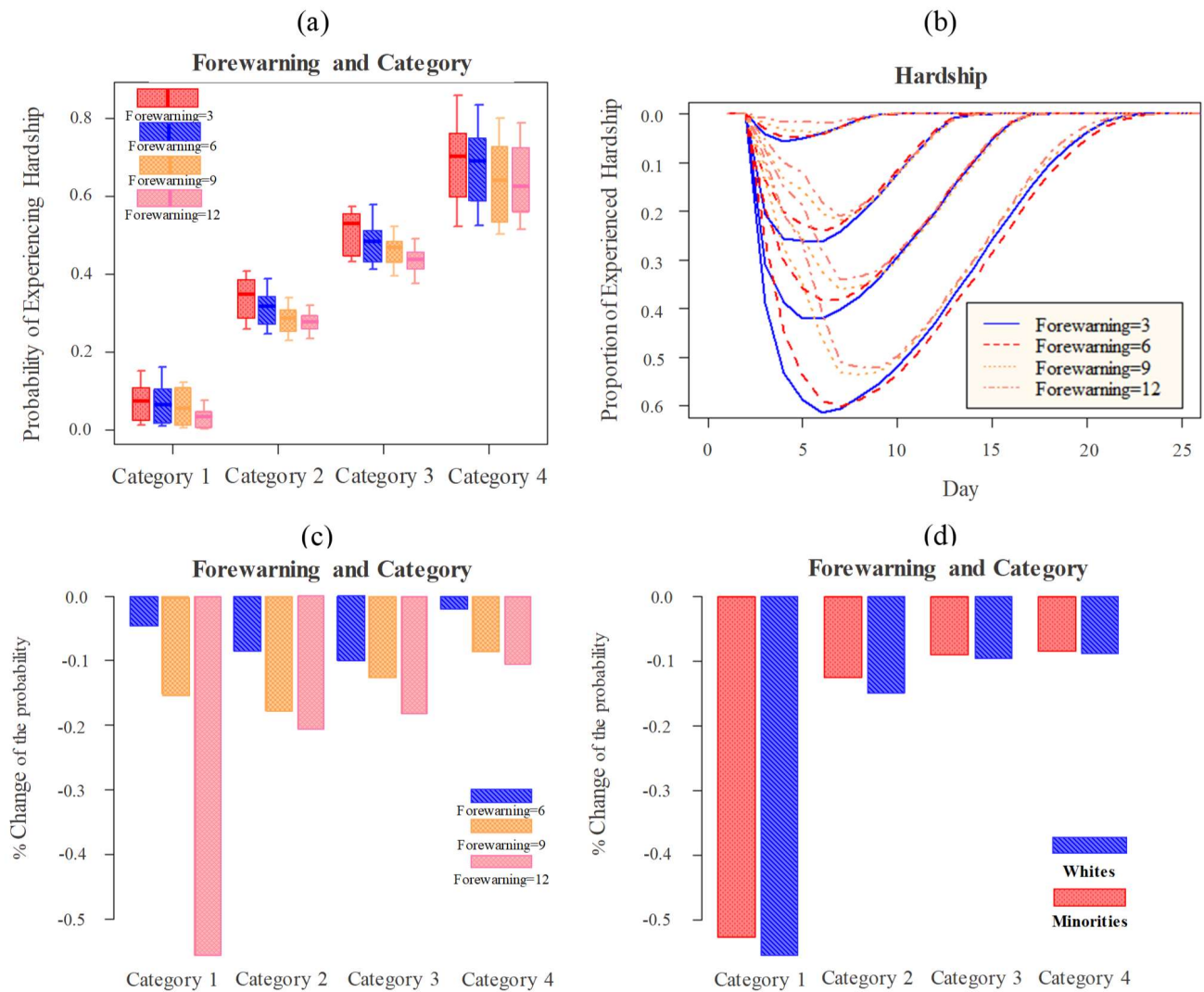

**FIGURE 17.** Effect of providing longer forewarning on the societal impacts of power outages under various hurricane intensities. (a) histograms of the probability of expiring hardship for each scenario. (b) average daily experienced hardship for each scenario. (c) percentage change of the reduction in the probability of experiencing hardship for each scenario compared to the forewarning equal to 3 days, and (d) percentage difference of the probability of experiencing hardship for the racial groups under each scenario.

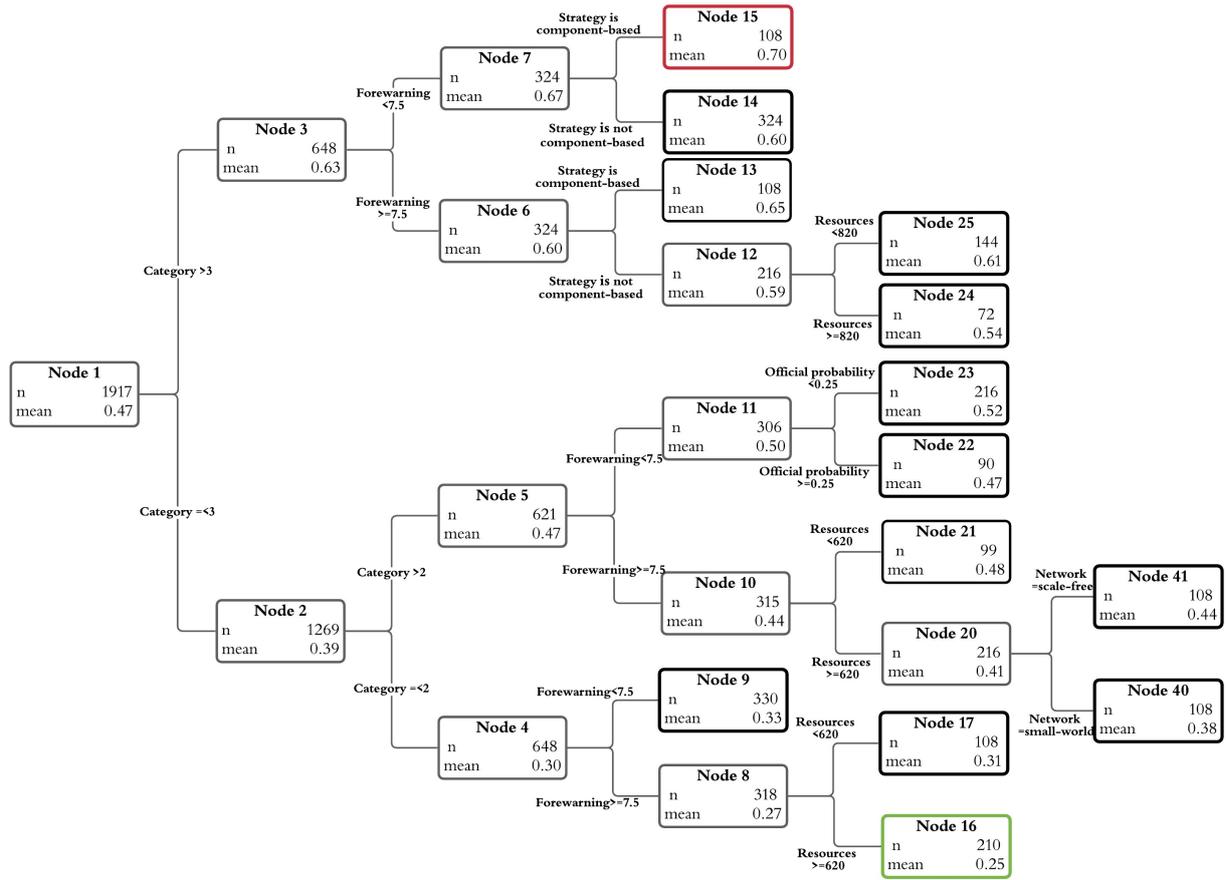

**FIGURE 18.** Classification and regression tree (CART) analysis for analyzing the effect of various strategies in reducing the societal impacts.

In this analysis, in addition to the described strategies for reducing the societal impacts (restoration activity, longer forewarning, and social network structure), also included are hurricane category, the number of restoration resources, and the information sharing probability of the officials are also included. The hurricane category has the greatest impact on household experienced hardship. A longer forewarning duration seems to have a great impact on reducing the societal impacts of the power outages. This pattern is consistent for different hurricane categories, which supports the suggestion that providing a longer forewarning could effectively reduce societal impacts. The effect of the restoration strategy and increasing the number of resources varies depending on the hurricane intensity. Improving the restoration strategy to focus on the needs of the population (population-based and SVI-based) seems to more effectively reduce societal impacts than increasing the number of resources in response to high-intensity hurricanes. The effect of increasing the number of resources, however, seems to be an effective approach for lower-severity events. Lastly, when considering the effect of longer forewarning and information sharing by the officials, the effect of the social network structure seems to be insignificant in reducing the societal impacts of disaster-induced prolonged power outages. The results show that hardships due to power outages during high-intensity hurricanes would be inevitable for minorities and other vulnerable populations unless two changes are implemented: power infrastructure systems should be strengthened to reduce their likelihood of failure and sufficient resources, focusing on socially vulnerable populations, should be earmarked for prioritizing power restoration.

## 7 CONCLUDING REMARKS

This study presents a new computational simulation framework for modeling the complex hazard-human-infrastructure nexus to better integrate social equity considerations into resilience assessments. The proposed integrated multi-agent simulation model enables capturing of the complex interactions between hazard, risk and restoration process, and households' decision-making behaviors. This new computational model enables consideration of heterogeneity in the impact of infrastructure service disruption in affected communities.

The model enables a combined evaluation of the effects of hazard characteristics, population attributes and decision-making processes, and physical infrastructure network topology and vulnerability in facilitating more equitable resilience assessments. While the current literature reports studies of various computational models for assessing infrastructure resilience, the majority of existing models, focusing primarily on physical systems, fail to consider a population's interactions with these systems and their services during disasters. The proposed computational framework captures and models the underlying dynamic mechanisms and complex interactions among hazard, physical networks, and household behavior. Thus, this paper contributes to the field of computer-aided infrastructure engineering by (1) abstracting the complex mechanisms that lead to the societal impacts of hurricane-induced power outages; (2) simulating societal impacts by using theoretical models and empirical data and capturing and modeling the interactions between hazard, power network, and households' behavior; and (3) devising an approach to meet the need for equitable resilience assessment in infrastructure systems. The multi-agent simulation model enables the inclusion of the social dimension in the resilience assessment of the infrastructure system. The model capabilities enable assessment as to what extent different strategies moderate the impacts for each segment of the community. The output results would inform about overall societal impact on the community and the distributional impact on the various segments of the community. By enabling decision-makers to conduct scenario analysis of strategies for reducing societal impacts of power outages by examining the effects of changing variables, such as restoration strategies, forewarning time, and household social network structure, the model enables decision-makers to reduce overall societal impacts. The proposed model could be used by emergency and infrastructure managers and operators to better prioritize resource allocation to their hazard mitigation investments and restorations to reduce the societal impacts of infrastructure disruptions. In addition, the integrated simulation framework that captures the complex interactions among hazard characteristics, population behaviors, and physical infrastructure network properties could provide a tool and simulated data for developing more interdisciplinary disaster resilience theories and examining complex phenomena, which could not be evaluated using empirical and observational data (Mostafavi and Ganapati 2019).

# APPENDIX A

# Pseudo Algorithms

**TABLE A-1.** Pseudo algorithms for the damage from the hurricanes based on the fragility equations.

```
Algorithm 1 fragility curve
input: probability of failure for each element
output: element that fails
 1: function FRAG CURVE(PoF)
 2:     for day d in hurricane duration do
 3:         for element e in agents do
 4:             if e does not fail then
 5:                 R(Random) ← windspeed
 6:                 if R < PoF then
 7:                     e fails and remove all connection link to e
 8:                 end if
 9:             end if
10:         end for
11:     end for
12: end function
```

**TABLE A-2.** Pseudo algorithms for the damage from the cascading effect.

```
Algorithm 2 cascade failure
 1: function SUBSTATION FAIL(PoF(Substation), network(ArrayList))
 2:     if Substation s fails then
 3:         for transmission t connected to s do
 4:             t fails
 5:         end for
 6:     end if
 7: end function
 8: function TRANSMISSION FAIL(PoF(transmission), network(ArrayList))
 9:     if transmissions sall connected to Substation s fail then
10:         s fails
11:         call substation fail
12:     end if
13: end function
```

**TABLE A-3.** Pseudo Households decision making and protective action.

```
Algorithm 3 preparation
 1: function PREPARATION(demographic features, CumulativeLogi formula, prepare lambda)
 2:     for d in information propagation period do
 3:         pre1 − 5 ← CumulativeLogic(demographic)
 4:         PortionPrepare ← NeighborPrepareSum/neighborSize
 5:         r ← Random + PortionPrepare ∗ preparelambda
 6:         if r < pre1 then
 7:             prepare ← 1
 8:         else if R < pre2 then
 9:             prepare ← 2
10:         else if R < pre3 then
11:             prepare ← 3
12:         else if R < pre4 then
13:             prepare ← 4
14:         else
15:             prepare ← 5
16:         end if
17:     end for
```

**TABLE A-4.** Pseudo algorithms for the restoration activity and prioritization.

```
Algorithm 4 restoration(use population based scenario)
input: Strategy && resource
 1: function RESTORATION(Strategy scenario,origin
    resource r,resource increase speed a,number of
    failedtracts t)
 2:     for h in resource increasing period do
 3:         r ← r + a
 4:     end for
 5:     while r > 0 and fix tracts < t do
 6:         if r > xr then
 7:             fix tract[x]
 8:             r ← r - xr
 9:             fixtracts ← fixtracts + 1
10:         else
11:             continue
12:         end if
13:     end while
14:     for Pole p in tract[x] do
15:         if p.nearest substaion(s) not fixed and r > sr
    then
16:             fix s
17:             r ← r - sr
18:         end if
19:     end for
20:     for transmission t in tract[x] do
21:         if t not fixed and r > tr then
22:             fix t
23:             r ← r - tr
24:         end if
25:     end for
26:     for Pole p in tract[x] do
27:         if p not fixed and r > pr then
28:             fix p
29:             r ← r - pr
30:         end if
31:     end for
32:     for Substation s in tract[x] do
33:         if s not fixed and r > sr then
34:             fix s
35:             r ← r - sr
36:         end if
37:     end for
38: end function
```

# APPENDIX B: MODEL DEVELOPMENT

## Fragility Curves and Restoration Resources

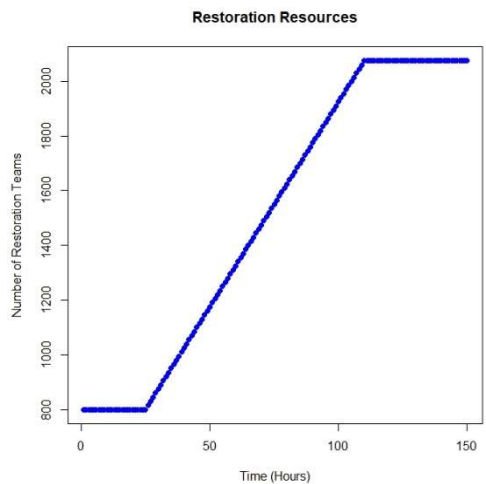

**FIGURE B-1.** Number of added resources for the restoration activity.

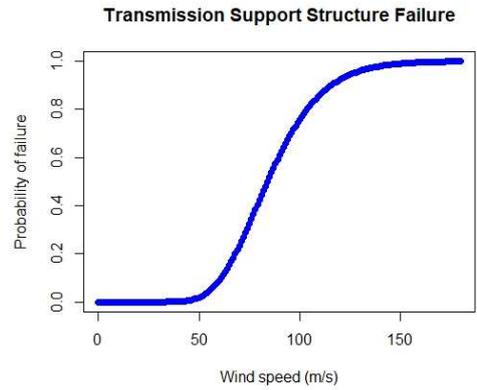

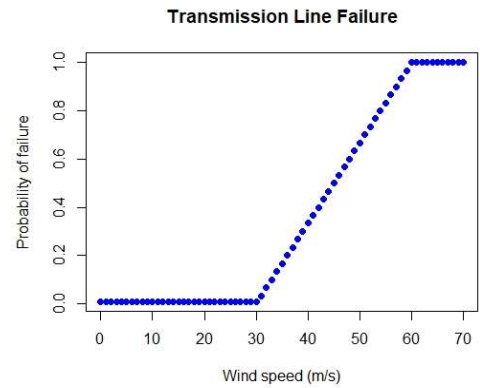

**FIGURE B-2.** Transmission distribution network fragility curve.

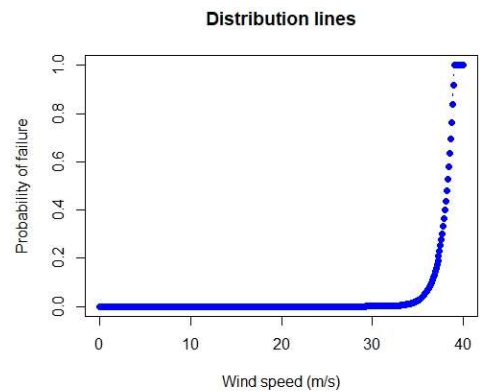

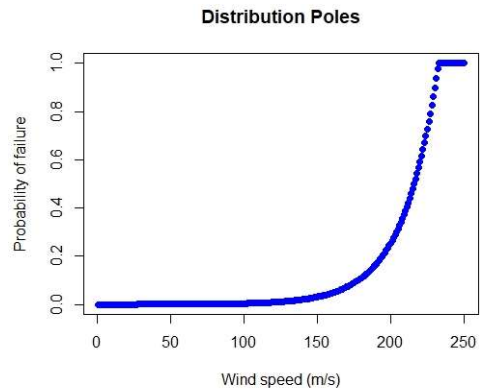

**FIGURE B-3.** Distribution network fragility curve.

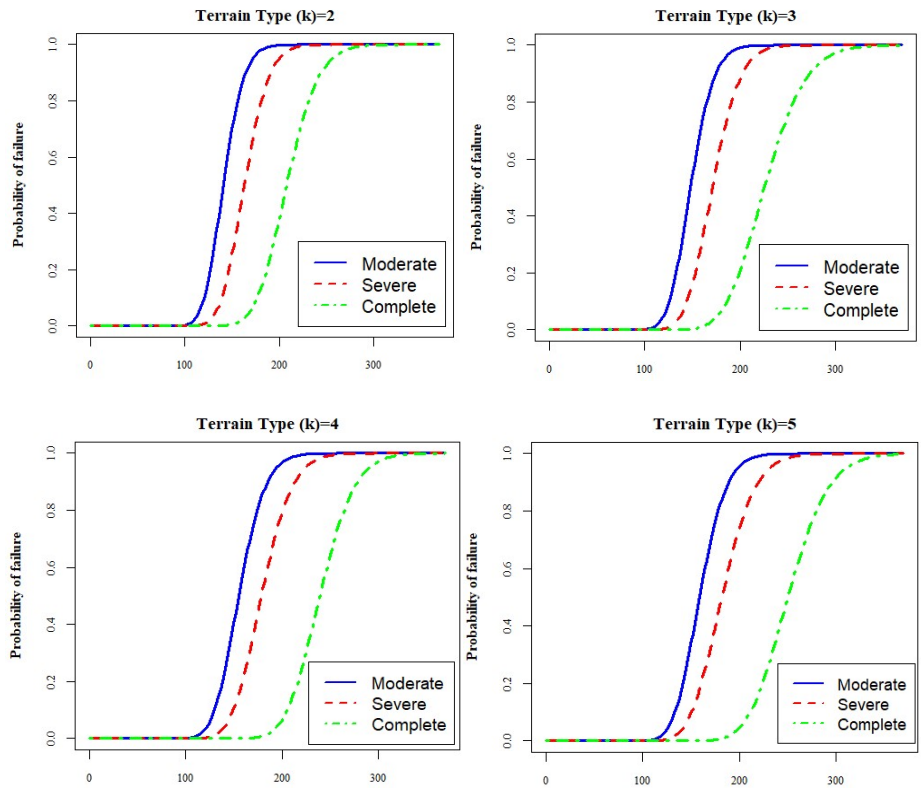

**FIGURE B-4.** Substitution fragility curve.

**TABLE B-1.** Required resources for the damage to each component.

| Damaged component | Restoration Time | Needed resources |
|---|---|---|
| Load Substations | Moderate: $N^*(72h, 36h)$, Severe: $N(168h, 84h)$ and Complete: $N(720h, 360h)$ | 6 14 60 |
| Transmission Towers | $N(72h, 36h)$ | 6 |
| Transmission Lines | $N(48h, 24h)$ | 4 |
| Distribution Poles | $N(10h, 5h)$ | 1 |
| Distribution Lines | $N(8h, 4h)$ | 1 |

* Note: $N(a,b)$ refers to the randomly generated number from a normal distribution with mean=$a$ and standard deviation=$b$ (Mensah 2015)

## Household Agents

### Model description

In these models, the zone of tolerance would be calculated through the process and depending on the three variables. The households' zone of tolerance is a function of the household's need, substitute, and preparedness level. The following equation describes the relationships among the variables:

$$\mu = \exp[1.7762 - 0.5130x_s + 0.1827x_n + 0.2664x_p]$$

Therefore, in this model, we needed to calculate the three factors of substitute, need, and preparedness.

### Need

The needed variable is inherent based on the socio-demographic characteristics of the household. The table below shows the influencing variables:

**TABLE B-2.** Influencing factors of the households' need.

| Variable | Measure |
|---|---|
| Race minority | "Yes" =1, "No" =2 |
| Mobility issue | "Yes"=1, "No"=2 |
| Young children (age-10) | "Yes"=1, "No"=2 |
| Medical | "Yes"=1, "No"=2 |

The variables in the model are socio-demographic characteristics; therefore, we implemented a simulated sample of the population for determining these variables.

The cumulative logit models with proportional odds were used for modeling the parameter; here, there are four intercepts, which means there exist four equations for calculating the probability of the five need levels. The general equation for this model is as follows:

$$logit[P(Y \leq j)] = \log\left[\frac{P(Y \leq j)}{1 - P(Y \leq j)}\right]$$
$$= \log\left[\frac{\pi_1 + \cdots + \pi_j}{\pi_{j+1} + \cdots + \pi_J}\right], \quad j = 1, \ldots, J-1$$

Here, instead of directly calculating the probability of each level (for example, the probability of need to be 1 ($p(y = 1)$)), we will calculate the $p(Y \leq 1)$. But $P(y \leq 1) = P(y = 1)$; thus, we can calculate the probability of the first level, p(y=1), by the following equation:

$$\log \frac{p(y=1)}{1-p(y=1)} = 0.44441 + 0.89646x_r - 0.51914x_m + 0.21971x_a - 0.30319x_m$$

Then the probability of ($p(y = 1)$) would be determined based on the following equation:

$$p(y = 1) = \frac{e^{[p(y=1)]}}{1 + e^{[p(y=1)]}}$$

Then, the next probability would be the probability of $p(Y \leq 2)$, which is $P1+P2$. Therefore, we can calculate the probability of the second one based on the difference between this probability and the one calculated in the previous step:

$$\log \frac{p(y1) + p(y2)}{p(y3) + p(y4) + p(y5)} = 1.79242 + 0.89646x_r - 0.51914x_m + 0.21971x_a - 0.30319x_m$$

Therefore, $p(y \leq 2)$ would be calculated based on the following equation:

$$p(y \leq 2) = \frac{e^{[p(y \leq 2)]}}{1 + e^{[y \leq 2]}}$$

Thus $p(2)$ would be the difference between the two probabilities. This will be continued until we have used the third and fourth intercepts. Lastly, the probability of the final level p5 would be calculated by $1-p(y \leq 4)$. Here, $p(y \leq 4)$ is equal to the last equation using intercept 4.

TABLE B-3. Influencing factors of the households' need.

| Variable | Estimate | P-value |
|---|---|---|
| (Intercept):1 | 0.444 | 0.125 |
| (Intercept):2 | 1.792 | <.001 |
| (Intercept):3 | 3.344 | <.001 |
| (Intercept):4 | 4.992 | <.001 |
| Racial minority | 0.896 | <.001 |
| Mobility issue | -0.519 | <.001 |
| Having children (<10) | 0.220 | 0.050 |
| Medical issue | -0.303 | <.001 |

## Substitute

We calculate the probability of getting a generator by using logistic regression. We calculate the probability of getting a generator by using logistic regression. Here the probability depends on the income, self-efficacy, ownership, and the household's expectations of the disruptions. The table below shows the variables:

TABLE B-4. Influencing factors of the households' protective action (buying a generator).

| Variable | Measure |
|---|---|
| Income | "Less than $25,000" = 1, "$25,000 - $49,999" = 2,"$50,000 - $74,999"=3, "$75,000 - $99,999" = 4, "$100,000 - $124,999"=5,"$125,000 - $149,999"=6," More than $150,000"=7 |
| Expectations | The number calculated in the previous step |
| Ownership | "Renter" (1), "Owner" (0) |
| Self-efficacy | "Strongly low"=1, "Somewhat low"=2, "Medium"=3, "Somewhat high"=4, "Strongly high"=5 |

Logistic regression relates the predictors to the logit based on the following equation:

$$P_s = \log \frac{p(y=1)}{1-p(y=1)} = -2.53950 + 0.07416x_i - 0.93270x_o + 0.48647\log(x_e + 1) + 0.26128x_{se}$$

Here, the log transformation was conducted on the expectation variable. Then the probability of having a generator or $p(y = 1)$ would be determined based on the following equation:

$$p(y = 1) = \frac{e^{[P_s]}}{1 + e^{[P_s]}}$$

## Preparation

This variable was modeled in a similar fashion as the substitute. The main variable which makes it a process variable is the forewarning. This variable depends on the following factors: having a vehicle, previous experience, being elderly, ownership, forewarning, distance to the supermarket, and self-efficacy. We calculated the probability of preparedness by using logistic regression. The table below shows the variables:

TABLE B-5. Influencing factors of the households' preparation.

| Variable | Measure |
|---|---|
| Vehicle vulnerability | "Did not have a car" =1, "I have it" =0 |
| Experience | The number calculated in the previous step |
| Ownership | Renter (1), Owner (0) |
| Self-efficacy | "Strongly low"=1, "Somewhat low"=2, "Medium"=3, "Somewhat high"=4, "Strongly high"=5 |
| Elderly | Yes (1), No (0) |
| Forewarning | Number of days |
| Distant to supermarket | miles |

Note: Distance was simulated from a normal distribution with mean 5 and variance 30.

Logistic regression relates the predictors to the logit based on the following equation:

$$P_p = \log \frac{p(y=1)}{1-p(y=1)} = 1.89292 - 0.58174 x_v \\ - 1.11299 x_e + 0.44445 x_{el} - 0.60578 x_o \\ + 0.08802 x_f - 0.02362 x_d + 0.50834 x_{se}$$

Then the probability of having a generator or $p(y=1)$ would be determined based on the following equation:

$$p(y=1) = \frac{e^{[P_p]}}{1+e^{[P_p]}}$$

## Self-efficacy

This variable defines to what extent the households believe in the effectiveness of the preparedness actions. The table below shows the influencing variables: Ownership, having social capital, having a chronic disease, and a medical condition.

**TABLE B-6.** Influencing factors of the households' level of self-efficacy.

| Variable | Measure |
|---|---|
| Ownership | Yes (1), No (0) |
| Social capital | Yes (1), No (0) |
| Chronic disease | Yes (1), No (0) |
| Medical | Yes (1), No (0) |

The calculation of the probabilities for each level should be done using the procedure explained in the need section.

**TABLE B-7.** Influencing factors of the households' level of self-efficacy.

| Variable | Estimate | P-value |
|---|---|---|
| (Intercept):1 | -3.191 | <.001 |
| (Intercept):2 | -1.792 | <.001 |
| (Intercept):3 | -0.551 | 0.009 |
| (Intercept):4 | 1.458 | <.001 |
| Ownership | 0.339 | <.001 |
| Medical | -0.245 | 0.016 |
| Chronic disease | -0.237 | 0.029 |
| Social capital | 0.217 | <.04 |

## Experience

This variable is calculated to find those with previous disaster experience. Having previous experience with a disaster depends on the duration of the time they have lived in their state, racial minority, elderly, and having a child.

**TABLE B-8.** Influencing factors of the households' level of experience.

| Variable | Measure |
|---|---|
| Having a child (Age-10) | Yes (1), No (0) |
| Race | Yes (1), No (0) |
| State duration | Number of years |
| Elderly | Yes (1), No (0) |

State duration should be simulated based on a normal distribution and mean 25 and standard deviation 15 (variance of 225). Logistic regression relates the predictors to the logit based on the following equation:

$$\log \frac{p(y=1)}{1-p(y=1)} = 1.371844 + 0.020162 x_{sd} \\ - 0.656271\, x_r - 0.366558 x_a \\ + 0.272127 x_e$$

Then the probability of having a generator or $p(y=1)$ would be determined based on the following equation:

$$p(y=1) \\ = \frac{e^{[-1.98711+0.1245\,i-0.71779\,x_o+0.37576\log(x_e+1)]}}{1+e^{[-1.98711+0.12456 x_i-0.71779\,x_o+0.37576\log(x_e+1)]}}$$

## ACKNOWLEDGMENT

The authors would like to acknowledge the funding support from the National Science Foundation under grant number 1846069 and National Academies' Gulf Research Program Early-Career Research Fellowship. Any opinions, findings, conclusions, or recommendations expressed in this research are those of the authors and do not necessarily reflect the view of the funding agencies.

ESMALIAN ET AL | 23Chakalian, P. M., Asce, S. M., Kurtz, L. C., Asce, S. M., Hondula, D. M., and Ph, D. (2019). "After the Lights Go Out : Household Resilience to Electrical Grid Failure Following Hurricane Irma." 20(September 2017), 1–14.

Coleman, N., Esmalian, A., and Mostafavi, A. (2019). "Equitable Resilience in Infrastructure Systems: Empirical Assessment of Disparities in Hardship Experiences of Vulnerable Populations during Service Disruptions." *Natural Hazards Review*, 21(24), 04020034.

Coleman, N., Esmalian, A., and Mostafavi, A. (2020). "Anatomy of Susceptibility for Shelter-in-Place Households Facing Infrastructure Service Disruptions Caused by Natural Hazards." *International Journal of Disaster Risk Reduction*, 50, 101875.

Cremen, G., and Galasso, C. (2021). "A decision-making methodology for risk-informed earthquake early warning." *Computer-Aided Civil and Infrastructure Engineering*, 1–15.

Dai, Q., Zhu, X., Zhuo, L., Han, D., Liu, Z., and Zhang, S. (2020). "A hazard-human coupled model (HazardCM) to assess city dynamic exposure to rainfall-triggered natural hazards." *Environmental Modelling and Software*, Elsevier Ltd, 127(March), 104684.

Dale, C. J. (1985). "Application of the proportional hazards model in the reliability field." *Reliability Engineering*, 10(1), 1–14.

Duffey, R. B. (2019). "Power Restoration Prediction Following Extreme Events and Disasters." *International Journal of Disaster Risk Science*, Beijing Normal University Press, 10(1), 134–148.

Edison Electric Institute. (2016). *Understanding the Electric Power Industry's Response and Restoration Process The Storm Process The Restoration Storm Restoration Process The Storm Restoration Process*.

Eid, M. S., and El-adaway, I. H. (2018). "Decision-Making Framework for Holistic Sustainable Disaster Recovery: Agent-Based Approach for Decreasing Vulnerabilities of the Associated Communities." *Journal of Infrastructure Systems*, 24(3), 04018009.

Esmalian, A., Dong, K., and Mostafavi, A. (2020a). "Empirical Assessment of Household Susceptibility to Hazards-induced Prolonged Power Outages." *Construction Research Congress*, 933–941.

Esmalian, A., Dong, S., Coleman, N., and Mostafavi, A. (2021). "Determinants of risk disparity due to infrastructure service losses in disasters: a household service gap model." *Risk Analysis*, 0(0).

Esmalian, A., Dong, S., and Mostafavi, A. (2020b). "Susceptibility Curves for Humans: Empirical Survival Models for Determining Household-level Disturbances from Hazards-induced Infrastructure Service Disruptions." *Sustainable Cities and Society*, 66, 102694.

FEMA. (2008). "Hazards U.S. multi-hazard (HAZUS-MH) assessment tool vs 1.4." Washington, DC.

Figueroa-candia, M., Felder, F. A., and Coit, D. W. (2018). "Resiliency-based optimization of restoration policies for electric power distribution systems." *Electric Power Systems Research*, Elsevier B.V., 161, 188–198.

Flanagan, B. E., Gregory, E. W., Hallisey, E. J., Heitgerd, J. L., and Lewis, B. (2011). "A Social Vulnerability Index for Disaster Management." *Journal of Homeland Security and Emergency Management*, 8(1).

Gegner, K. M., Birchfield, A. B., Xu, T., Shetye, K. S., and Overbye, T. J. (2016). "A methodology for the creation of geographically realistic synthetic power flow models." *2016 IEEE Power and Energy Conference at Illinois, PECI 2016*, 1–6.

Gori, A., Gidaris, I., Elliott, J. R., Padgett, J., Loughran, K., Bedient, P., Panakkal, P., and Juan, A. (2020). "Accessibility and Recovery Assessment of Houston's Roadway Network due to Fluvial Flooding during Hurricane Harvey." *Natural Hazards Review*, 21(2), 04020005.

Guidotti, R., Gardoni, P., and Rosenheim, N. (2019). "Integration of physical infrastructure and social systems in communities' reliability and resilience analysis." *Reliability Engineering and System Safety*, Elsevier Ltd, 185(February 2018), 476–492.

Guikema, S. D., Nateghi, R., Quiring, S. M., Staid, A., Reilly, A. C., and Gao, M. (2014). "Predicting Hurricane Power Outages to Support Storm Response Planning." *IEEE Access*, 2, 1364–1373.

Haer, T., Botzen, W. J. W., and Aerts, J. C. J. H. (2016). "The effectiveness of flood risk communication strategies and the influence of social networks-Insights from an agent-based model." *Environmental Science and Policy*, Elsevier Ltd, 60, 44–52.

Haer, T., Botzen, W. J. W., Moel, H. De, and Aerts, J. C. J. H. (2017). "Integrating Household Risk Mitigation Behavior in Flood Risk Analysis : An Agent-Based Model Approach." 37(10).

Hahn, G. J. (1972). "Sample Sizes for Monte Carlo Simulation." *IEEE Transactions on Systems, Man and Cybernetics*, SMC-2(5), 678–680.

Hassan, E. M., and Mahmoud, H. (2021). "Healthcare and education networks interaction as an indicator of social services stability following natural disasters." *Scientific Reports*, Nature Publishing Group UK, 11(1), 1–15.

Holmgren, A. J. (2006). "Using Graph Models to Analyze the Vulnerability of Electric Power Networks." 26(4).

Horney, J. (2008). "Factors associated with hurricane preparedness: Results of a pre-hurricane assessment." *Journal of Disaster Research*, 3(2).

Horney, J., Snider, C., Malone, S., and Cross, R. (2014). "Factors Associated with Hurricane Preparedness : Results of a Pre-Hurricane Assessment." (November 2007).

Kashani, H., Movahedi, A., and Morshedi, M. A. (2019). "An agent-based simulation model to evaluate the response to seismic retrofit promotion policies." *International Journal of Disaster Risk Reduction*, Elsevier Ltd, 33(March 2018), 181–195.

Lindell, M. K., and Hwang, S. N. (2008). "Households' perceived personal risk and responses in a multihazard environment." *Risk Analysis*, 28(2), 539–556.

Liu, C., Ouyang, M., Wang, N., Mao, Z., and Xu, X. (2021). "A heuristic method to identify optimum seismic retrofit strategies for critical infrastructure systems." *Computer-Aided Civil and Infrastructure Engineering*, 1–17.

Liu, H., Davidson, R. A., and Apanasovich, T. V. (2007). "Statistical forecasting of electric power restoration times in hurricanes and ice storms." *IEEE Transactions on Power Systems*, 22(4), 2270–2279.

Ma, L., Christou, V., and Bocchini, P. (2019). "Probabilistic simulation of power transmission systems affected by hurricane events based on fragility and AC power flow analyses." *13th International Conference on Applications of Statistics and Probability in Civil Engineering, ICASP 2019*, (2014).

Mensah, A. F. (2015). "Resilience Assessment of Electric Grids and Distributed Wind Generation under Hurricane Hazards." Rice university.

Mensah, A. F., and Dueñas-Osorio, L. (2016). "Efficient Resilience Assessment Framework for Electric Power Systems Affected by Hurricane Events." *Journal of Structural Engineering (United States)*, 142(8), 1–10.

Mitsova, D., Esnard, A. M., Sapat, A., and Lai, B. S. (2018). "Socioeconomic vulnerability and electric power restoration timelines in Florida: the case of Hurricane Irma." *Natural Hazards*, Springer Netherlands, 94(2), 689–709.

Mitsova, D., Esnard, A., Sapat, A., Lamadrid, A., Escaleras, M., and Velarde-perez, C. (2021). "Effects of Infrastructure Service Disruptions Following Hurricane Irma : Multilevel Analysis of Postdisaster Recovery Outcomes." 22(1), 1–15.

Morss, R. E., Demuth, J. L., Lazrus, H., and Palen, L. (2017). "Hazardous Weather Prediction and Communication in the Modern Information Environment." (June 2018).

Morss, R. E., Mulder, K. J., Lazo, J. K., and Demuth, J. L. (2016). "How do people perceive, understand, and anticipate responding to flash flood risks and warnings? Results from a public survey in Boulder, Colorado, USA." *Journal of Hydrology*, Elsevier B.V., 541, 649–664.

Mostafavi, A. (2018). "A system-of-systems framework for exploratory analysis of climate change impacts on civil infrastructure resilience." *Sustainable and Resilient Infrastructure*, Taylor & Francis, 3(4), 175–192.

Mostafavi, A., Abraham, D., DeLaurentis, D., Sinfield, J., Kandil, A., and Queiroz, C. (2016). "Agent-Based Simulation Model for Assessment of Financing Scenarios in Highway Transportation Infrastructure Systems." *Journal of Computing in Civil Engineering*, 30(2), 04015012.

Mostafavi, A., and Ganapati, N. E. (2019). "Toward Convergence Disaster Research: Building Integrative Theories Using Simulation." *Risk Analysis*, n/a(n/a).

Murray, K., and Bell, K. R. W. (2014). "Wind related faults on the GB transmission network." *2014 International Conference on Probabilistic Methods Applied to Power Systems, PMAPS 2014 - Conference Proceedings*, IEEE, 1–6.